\documentclass[]{emulateapj}
\usepackage{amsmath}
\usepackage{amssymb}
\usepackage{epstopdf}

\slugcomment{Accepted for publication in \apj}
\shorttitle{The Chemical Imprint of Silicate Dust on the Most Metal-Poor Stars}
\shortauthors{Ji, Frebel, \& Bromm}

\begin{document}
\title{The Chemical Imprint of Silicate Dust on the Most Metal-Poor Stars} 
\author{Alexander P. Ji\altaffilmark{1}}
\author{Anna Frebel\altaffilmark{1}}
\author{Volker Bromm\altaffilmark{2}}
\altaffiltext{1}{Kavli Institute for Astrophysics and Space Research
  and Department of Physics, Massachusetts Institute of Technology, 77
  Massachusetts Avenue, Cambridge, MA 02139, USA; alexji@mit.edu,
  afrebel@mit.edu}
\altaffiltext{2}{Department of Astronomy and Texas Cosmology Center,
  University of Texas at Austin, 2511 Speedway, Austin, TX 78712-0259,
  USA; vbromm@astro.as.utexas.edu}

\begin{abstract}
We investigate the impact of dust-induced gas fragmentation on the
formation of the first low-mass, metal-poor stars ($<1M_\odot$) in the
early universe. Previous work has shown the existence of a critical
dust-to-gas ratio, below which dust thermal cooling cannot cause
gas fragmentation. 
Assuming the first dust is silicon-based, we compute
critical dust-to-gas ratios and associated critical silicon
abundances ($\mbox{[Si/H]}_{\text{crit}}$).
At the density and temperature associated with protostellar disks, we
find that a standard Milky Way grain size distribution gives
$\mbox{[Si/H]}_{\text{crit}} = -4.5 \pm 0.1$, while smaller grain sizes
created in a supernova reverse shock give
$\mbox{[Si/H]}_{\text{crit}} = -5.3 \pm 0.1$.
Other environments are not dense enough to be influenced by dust
cooling.
We test the silicate dust cooling theory by comparing to silicon
abundances observed in the most iron-poor stars ($\mbox{[Fe/H]}<-4.0$).
Several stars have silicon abundances low enough to rule out
dust-induced gas fragmentation with a standard grain size
distribution. Moreover, two of these stars have such low silicon
abundances that even dust with a shocked grain size distribution
cannot explain their formation. Adding small amounts of
carbon dust does not significantly change these conclusions.
Additionally, we find that these stars exhibit either high carbon with
low silicon abundances or the reverse.
A silicate dust scenario thus suggests that the earliest low-mass star
formation in the most metal-poor regime may have proceeded through two
distinct cooling pathways: fine structure line cooling and
dust cooling. This naturally explains both the carbon-rich and
carbon-normal stars at extremely low [Fe/H].
\end{abstract}


\section{Introduction} \label{section:introduction}
The formation of the first stars marks the beginnings of structure
formation, cosmic reionization, and chemical enrichment
(e.g., \citealt{Bromm09} and references within). These so-called
Population III stars formed out of metal-free primordial gas at the
centers  of dark matter minihalos
\citep{Couchman86,Haiman96,Tegmark97,Yoshida03}. Due to relatively
weak feedback and inefficient cooling, they had high characteristic
masses of order at least tens of solar masses and therefore short life
spans (e.g.,
\citealt{Abel02,Bromm02,Stacy10,Stacy12,Greif11,Hosokawa11}). 

Although the short lives of Population III stars implies that they
cannot be directly observed anymore, it is believed that the metals
released in their supernovae trigger a transition
from predominantly high mass star
formation to a low mass mode \citep{Bromm01,Schneider02}. The
chemical abundances of low-mass, metal-poor Population II stars in the
Milky Way stellar halo have been interpreted as traces of the
Population III star era (e.g., \citealt{Beers05},
\citealt{Frebel13a}). If this is indeed the case, then an understanding
of the formation process for Population II stars is one way to probe
the epoch of the first stars \citep{Tumlinson06,Karlsson13}.

However, unlike the formation of Population III stars, whose gas
properties and formation environments are relatively well-understood, 
the conditions for Population II star formation are quite uncertain
(e.g., \citealt{Bromm13}).
Introducing even trace amounts of metals significantly affects the
thermal behavior of collapsing gas clouds (e.g., \citealt{Omukai05}).
There are also many possible candidate environments that might be the
formation sites of these stars, ranging from the atomic cooling halos
of the first protogalaxies (e.g., \citealt{Wise07,Greif08,Greif10}) to
post-supernova shock regions (e.g., \citealt{Salvaterra04,Chiaki13b}). 
The two main theories for how metals cause low mass star formation
are gas cooling through atomic fine structure lines
\citep{Bromm03,Santoro06} and gas fragmentation induced by dust
continuum radiation (e.g., \citealt{Schneider06,Omukai10}).
We will refer to these as ``fine structure cooling'' and ``dust
cooling'', respectively. 

Fine structure cooling argues that in the absence of sufficient atomic metal
line cooling, gas clouds cannot quickly collapse beyond a ``loitering
state'' of $n \sim 10^4\,$cm$^{-3}$ and $T \sim 200\,$K
\citep{Bromm02}. The presence of
molecular hydrogen may smooth out this metallicity threshold (e.g.,
\citealt{Jappsen09a,Jappsen09b}), but only if there is no soft UV
Lyman-Werner (LW) background produced by the first stars,
capable of destroying molecular hydrogen \citep{Bromm01, SafShrad10}. 
Arguably, the presence of such a LW background is natural, as the same
stars that produced the first heavy elements would also emit LW
radiation; thus, the fine-structure threshold
is clearly imprinted, without H$_2$ cooling smoothing it out.
If the gas metallicity is above a critical metallicity of $Z/Z_\odot
\sim 10^{-3.5}$, the gas is unstable to vigorous fragmentation 
(e.g., \citealt{Santoro06,Smith09}).
The most important atomic species are carbon and oxygen
\citep{Bromm03}, so the theory predicts enhancements in these
elements. If correct, this is a natural explanation for the measured
carbon enhancement in many metal-poor stars \citep{Frebel07}. 
However, the Jeans mass of gas fragments formed by just fine
structure cooling is  $\geq 10M_\odot$, which is too massive for a
star formed early in the universe to survive until the present day
\citep{Klessen12}.

In contrast, dust cooling easily causes gas fragmentation at Jeans
masses of $\sim 0.1-1\, M_\odot$ because it becomes efficient only at
high gas densities and temperatures around $10^{12}$ cm$^{-3}$ and
$1000\,$K. The critical metallicity required for dust cooling to cause
fragmentation is also much lower at $Z/Z_\odot \sim 10^{-5}$
\citep{Omukai05,Tsuribe06,Schneider06,Clark08,Omukai10,Schneider12,Dopcke13}.
This dust must have been formed in early supernovae \citep{Gall11}.
Many dust models have been produced which turn supernova yields into
dust masses (e.g., \citealt{Todini01,Nozawa03,Schneider04,Bianchi07}).
Most of these models assume steady-state chemistry and use classical
nucleation theory to calculate dust yields. These approximations may
not applicable in a supernova outflow environment
\citep{Donn85,Cherchneff08,Cherchneff09,Cherchneff10}, although see
\citet{Paquette11} and \citet{Nozawa13}. Furthermore, significant
amounts of dust can also be destroyed in supernova reverse shocks
\citep{Silvia10}.

The large difference in the critical metallicity between these two
cooling mechanisms has sparked some debate about which one is most
relevant for the formation of low-mass metal-poor stars. 
This can be observationally tested, as the relevant cooling mechanisms
should leave an imprint on the observed chemical abundances.
\citet{Frebel07} observationally tested the fine structure cooling theory by
introducing the transition discriminant $D_{\text{trans}}$. They predicted
that metal-poor stars forming through this mechanism must have
$D_{\text{trans}} > -3.5 \pm 0.2$. Nearly all stars satisfy this criterion
(see \citealt{Frebel13a} for an updated $D_{\text{trans}}$ figure). The only
star known to violate the $D_{\text{trans}}$ criterion is SDSS~J1029151+1729 
\citep{Caffau11}. \citet{Schneider12b} and \citet{Klessen12} showed that dust cooling
was instead able to explain the formation of this star. More generally,
\citet{Schneider12} calculate a critical dust-to-gas ratio
($\mathcal{D}_{\text{crit}}$) that could in principle place an
observational restriction on dust cooling, similar to the
$D_{\text{trans}}$ restriction on fine structure cooling. However, for
metal-poor stars besides SDSS~J1029151+1729, the impact of dust
cooling has not been evaluated in detail.

Ideally, there would be general properties of supernova dust that
could be tested with observations of abundances in metal-poor stars.
Recently, \citet{Cherchneff10} have shown that when accounting for
non-equilibrium chemical kinetics in dust formation, dust yields are
significantly lower and dominated by silicon-based grains, rather than
the carbon grains that are typical results of most steady-state
models. There is some debate about the extent to which carbon
dust formation is suppressed (e.g., \citealt{Nozawa13}). However, if
indeed carbon dust formation is generally suppressed in the early
universe, the silicon abundance of metal-poor stars could be used as
an observational constraint on dust cooling processes.

In this paper, we investigate the impact that silicon-based dust could
have had on the formation process of the first low-mass stars.
Using the silicon-based dust compositions from \citet{Cherchneff10},
we compute critical silicon abundances and compare
them to observations of chemical abundances in long-lived metal-poor stars. In
Section~\ref{section:dustmodelintro}, we describe the dust models used for
this paper. In Section~\ref{section:critical}, we calculate critical
silicon abundances for our dust models, assessing how
differences in chemistry, grain size distribution, and environment affect
this critical threshold.
Our main results are found in
Section~\ref{section:data}, where we compare our critical silicon
abundances to measurements of metal-poor stars. 
Section~\ref{section:dustvsFS} considers evidence for two
distinct formation pathways of low-mass metal-poor stars, and
Section~\ref{section:DLA} discusses the potential for Damped
Lyman-$\alpha$ systems to help constrain the star formation
environments. After outlining important caveats in
Section~\ref{section:caveats} 
(particularly related to the production of carbon dust), 
we conclude in Section~\ref{section:conclusion}.

\section{Dust Models}\label{section:dustmodelintro}
We first present the dust models used in this paper in
Section~\ref{section:dustmodels}. We then discuss some processes in
these dust models which strongly inhibit carbon dust formation in
Section~\ref{section:whynocarbon}.

\subsection{Dust Chemical Composition and Size Distributions}\label{section:dustmodels}
We use the eight different silicon-based dust chemistries presented
in \citet{Cherchneff10}. We assume these are representative of typical
dust yields in the early universe.
The dust masses are given in Table~\ref{tbl:1}. Although the eight
different models represent different assumptions about the
nature of the supernovae and the dust condensation process, we
simply take them as plausible variations in the chemical composition
of dust. The dominant dust species are SiO$_2$, Mg$_2$SiO$_4$,
amorphous Si, and FeS.

For our calculations in Section~\ref{section:dustcooling}, we require
a dust grain size distribution. However, \citet{Cherchneff10}
do not compute grain size distributions for their dust models. We thus
consider two simple but well-motivated grain size distributions. The
first is a \citet{Pollack94} ``standard'' size distribution. This was
used in \citet{Omukai10}, and it is similar to the Milky Way grain
size distribution used in \citet{Dopcke13}. For spherical
dust grains of radius $a$:
\begin{equation} \label{eq:stdsizedistr}
  \frac{dn_{\text{standard}}}{da} \propto \begin{cases}
    1 & a < 0.005 \mu \text{m} \\
    a^{-3.5} & 0.005 \mu \text{m} < a < 1 \mu \text{m} \\
    a^{-5.5} & 1 \mu \text{m} < a < 5 \mu \text{m}
  \end{cases}
\end{equation}
We also consider a grain size distribution that approximates the
effect of running a post-supernova reverse shock through newly created
dust, based on the size distributions calculated in \citet{Bianchi07}:
\begin{equation} \label{eq:shocksizedistr}
  \frac{dn_{\text{shock}}}{da} \propto \begin{cases}
    1 & a < 0.005 \mu \text{m} \\
    a^{-5.5} & a > 0.005 \mu \text{m}
  \end{cases}
\end{equation}
From now on, we will refer to these two grain size distributions as
the ``standard'' and ``shock'' size distributions.
For simplicity, we assume that each type of dust grain has the same
grain size distribution,
though it may also be possible to calculate a
  good approximation to the grain size distribution using classical
  nucleation theory \citep{Paquette11}.
We normalize the size distributions to number
of particles per unit dust mass (cm$^{-1}$ g$^{-1}$) by using the
amount of dust mass formed and the solid-phase chemical density of
each type of dust \citep{Semenov03,HICC}. 

\subsection{Silicate or Carbon Dust?} \label{section:whynocarbon}
We use the \citet{Cherchneff10} dust models to establish a critical
silicon criterion (Section~\ref{section:sicrit}). Thus, our results
crucially depend on the assumption that the dust composition is
largely silicon-based.  The most significant non-silicate dust is
typically amorphous carbon. We thus briefly describe why carbon
dust formation is almost completely inhibited in these models. We
refer the reader to \citet{Cherchneff09,Cherchneff10} for a more
extensive discussion.

The chemical mechanisms that inhibit carbon dust formation depend on
the C/O ratio in the supernova ejecta.  When the C/O ratio is less
than one, CO formation rapidly depletes the available carbon. Although
there are processes that can destroy this supply of CO and form short
carbon chains, subsequent oxidation of these chains inhibits dust
formation. This effect is seen despite accounting for non-thermal
processes such as the destruction of CO through high energy Compton
electrons \citep{Cherchneff10}.  When the C/O ratio is greater than
one, small carbon clusters can form but are rapidly destroyed by the
He$^+$ ions that accompany large amounts of carbon.

In radial distributions of supernova ejecta, carbon is always
accompanied by large oxygen or helium abundances (e.g.,
\citealt{Nozawa03}). However, if the supernovae ejecta is poorly
mixed at a microscopic level, then carbon rich clouds may form
significant amounts of carbon dust in addition to silicate dust
\citep{Cherchneff10,Nozawa13}. 
Thus, the suppression of carbon dust may heavily depend on the level
of mixing, which itself depends on the details of the supernova
explosion.

We will follow the assumption of microscopically-mixed supernova
ejecta as in \citet{Cherchneff10}, which leads to silicate dust being
the dominant form of dust in the early universe.
A major motivation for investigating the consequences of silicon-based
dust is that silicon abundances measured from the most metal-poor
stars are comparable to the theoretical critical silicon abundances we
derive in Section~\ref{section:sicrit}, thus offering an empirical 
test of this important assumption.
For completeness, in Section~\ref{section:cdust}, we also explore the
impact that the formation of carbon dust would have on our results.

\section{Critical Silicon Abundance for Gas
  Fragmentation} \label{section:critical} 
In this section, we present the method for calculating the critical
silicon abundance ($\mbox{[Si/H]}_{\text{crit}}$) required for gas
fragmentation. We use a simplified model that only considers dust
thermal cooling and adiabatic compressional heating. Many previous
papers have studied these in detail, using a more comprehensive set of
cooling mechanisms that influence a large range of gas densities
(e.g., \citealt{Omukai00, Omukai05, Schneider06, Omukai10,
  Schneider12}). To derive the critical silicon abundance, we 
focus on the density regime where dust cooling dominates.

In Section~\ref{section:dustcooling} we show how we calculate the dust
cooling rate for a given dust model. In
Section~\ref{section:dcrit} we use the cooling rate to calculate a
critical dust-to-gas ratio \citep{Schneider12}, which we convert to a
critical silicon abundance in Section~\ref{section:sicrit}. 
In Section~\ref{section:environments}, we discuss
uncertainties in the Population II star forming environment and the
implications this may have for our critical silicon abundance.

\subsection{Calculating the Dust Cooling Rate} \label{section:dustcooling}
We describe how to calculate the gas cooling rate due to dust
emission. This calculation closely follows the method in
\citet{Schneider06}. For completeness and convenience of the reader,
we here give a brief summary.

Dust grain emission is well approximated by thermal radiation
\citep{Draine01}, in which case the cooling rate can be written 
\begin{equation} \label{eq:lambdad}
  \Lambda_{\text{d}} = 4 \sigma_{\text{SB}} T_{\text{d}}^4 \kappa_{\text{P}} \rho_{\text{d}} \beta_{\text{esc}}
\end{equation}
where $\sigma_{\text{SB}}$ is the Stefan-Boltzmann constant, $T_{\text{d}}$ is the dust
temperature, $\kappa_{\text{P}}$ is the temperature-dependent Planck
mean opacity of dust grains per unit \emph{dust} mass,
$\rho_{\text{d}}$ is the dust mass density, and 
$\beta_{\text{esc}}$ is the photon escape fraction. We define the
dust-to-gas ratio as 
\begin{equation} \label{eq:dusttogasratio}
  \mathcal{D} \equiv \rho_{\text{d}}/\rho
\end{equation}

\begin{figure}
\begin{center}
\includegraphics[width=9cm]{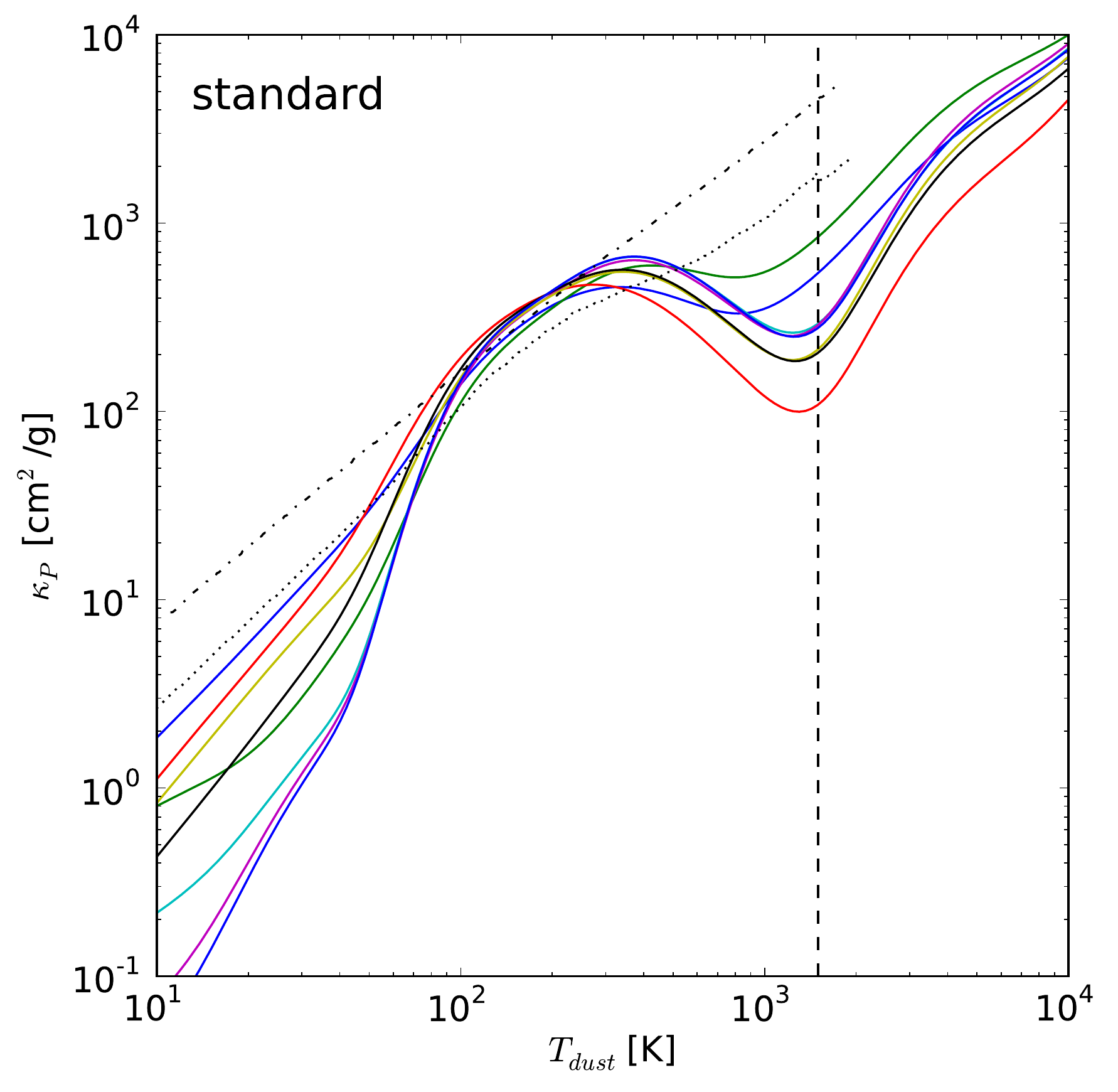} \\
\includegraphics[width=9cm]{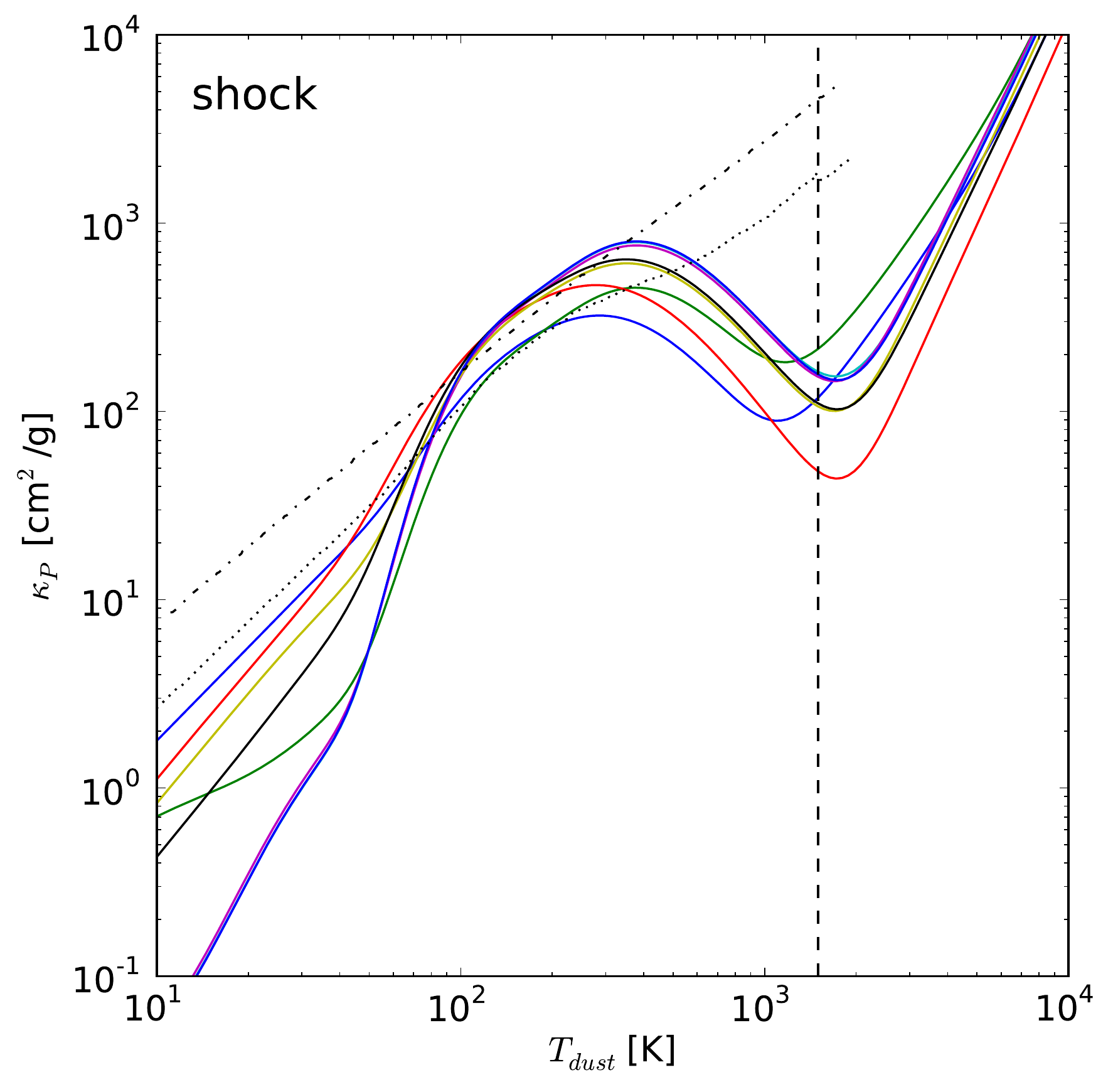}
\end{center}
\caption{Planck mean opacity. Top panel shows the standard size
  distribution, bottom panel the shock size
  distribution. The vertical dashed line indicates the dust
  sublimation temperatures of $1500\,$K. 
  For comparison, we also include Planck mean opacities for the
  core-collapse supernova model in \citet{Schneider06} (dotted line)
  and the metal-free supernova from \citet{Schneider12} (dash-dotted
  line), both of which contain significant amounts of carbon
  dust. These lines terminate when the dust has sublimated.
  \label{fig:kplanck}}
\end{figure}

For a given dust composition model, the Planck mean opacity is given
by 
\begin{equation} \label{eq:kappaplanck}
  \kappa_{\text{P}}(T_{\text{d}}) = \frac{\int_0^\infty\kappa_\lambda B_\lambda(T_{\text{d}})
    d\lambda}{\int_0^\infty B_\lambda(T_{\text{d}}) d\lambda}
\end{equation}
where $\kappa_\lambda$ is the wavelength-dependent opacity in
cm$^2$~g$^{-1}$ and $B_\lambda(T_{\text{d}})$ is the Planck specific
intensity. $\kappa_\lambda$ can be calculated by
\begin{equation} \label{eq:kappalambda}
  \kappa_\lambda = \sum_i f_i \kappa_\lambda^i \text{ with }
  \kappa_\lambda^i = \int_0^\infty Q_\lambda^i(a) \pi a^2
  \frac{dn^i}{da} da
\end{equation}
where $i$ denotes different dust species, $f_i$ is the mass fraction,
$Q_\lambda^i$ is the area-normalized absorption cross section, and
$dn^i/da$ is the size distribution. We calculate $Q_\lambda^i$ using
Mie theory, with optical constants taken from the sources listed in
Table~\ref{tbl:1}. If required, we linearly extrapolate the optical
constants on a log-log basis. 
We plot the Planck mean opacities for all our dust models in
Figure~\ref{fig:kplanck}, and for comparison we also include the 
Planck mean opacities for carbon-heavy dust models in
\citet{Schneider06} and \citet{Schneider12}. 

To calculate the dust temperature, we set the dust cooling
rate in Equation \ref{eq:lambdad} equal to the gas-dust collisional
heating rate \citep{Hollenbach79}: 
\begin{equation} \label{eq:energybalance}
  \Lambda_{\text{d}} = H_{\text{d}} = n n_{\text{d}} \sigma_{\text{d}}
  v_{\text{th}} f (2k_{\text{B}}T - 2k_{\text{B}}T_{\text{d}}) 
\end{equation}
where $n$ is the number density of atomic hydrogen, $n_{\text{d}}$ is
the number density of dust, $\sigma_{\text{d}}$ is the dust
geometrical cross section, $v_{\text{th}}$ is the thermal velocity of
atomic hydrogen, $f$ is a correction factor for species other than
atomic hydrogen, $T$ is the gas temperature, and $T_{\text{d}}$ is the
dust temperature. Note that the kinetic energy per colliding gas
particle is $2k_{\text{B}}T$ instead of $1.5k_{\text{B}}T$ because
higher energy particles collide more frequently \citep{DraineBook}.
We assume the gas has a Maxwellian velocity distribution so the
average velocity of atomic hydrogen is
\begin{equation}
  v_{\text{th}} = \left(\frac{8k_{\text{B}}T}{\pi m_{\text{p}}} \right)^{1/2}
\end{equation}
Since dust is most important at high gas densities, we assume the
hydrogen in the gas is fully molecular. Then neglecting the effects of
charge or sticking probabilities, we account for the differences in
number density and thermal velocity by setting $f = 1/2\sqrt{2} +
y_{\text{He}}/2$, where $y_{\text{He}} = n_{\text{He}}/n = 1/12$ for
primordial gas. We can also rewrite
\begin{equation}
  n_{\text{d}} \sigma_{\text{d}} = \rho_{\text{d}} S = \mathcal{D} \mu m_{\text{p}} n S 
\end{equation}
where $S$ is the total dust geometrical cross section per unit dust
mass defined by
\begin{equation}
  S = \sum_i f_i S_i \text{ with } S_i = \int_0^\infty\pi a^2
  \frac{dn^i}{da}da
\end{equation}
and $\mu = 1 + 4 y_{\text{He}} = 4/3$.

In general, solving Equation ~\ref{eq:energybalance} depends on
the dust-to-gas ratio $\mathcal{D}$ because the amount of dust may
influence $\beta_{\text{esc}}$. We assume $\beta_{\text{esc}} = \min
(1,\tau^{-2})$ which is suitable for radiative diffusion out of an optically
thick gas \citep{Omukai00}. The optical depth $\tau$ is given by:
\begin{equation} \label{eq:gasopacity} 
  \tau = (\kappa_{\text{gas}}\rho + \kappa_{\text{d}}\rho_{\text{d}}) \lambda_{\text{J}}
\end{equation}
$\kappa_{\text{gas}}$ is the continuum Planck mean opacity of primordial gas
from \citet{Mayer05}, $\kappa_{\text{d}}$ is the Planck mean opacity of
dust calculated in this paper, $\rho$ and $\rho_{\text{d}}$
are the densities of gas and dust respectively, and $\lambda_{\text{J}}$ is
the Jeans length. The Jeans length is the typical size of a dense core
of a uniformly collapsing spherical gas cloud (e.g.,
\citealt{Larson69}).
If the gas is optically thin ($\beta_{\text{esc}}=1$), it is possible
to solve for the dust temperature independently of
$\mathcal{D}$. However for the optically thick case, dust opacity will
affect the solution and cause some nonlinear dependence on $\mathcal{D}$.

If the dust temperature becomes too high, the dust will
sublimate. Different dust grains sublimate at different 
temperatures. We simplify this effect by assuming all grains sublimate
at $T_{\text{d}} = 1500\,$K, a typical temperature for non-carbon grains
\citep{Schneider06}. We set $\kappa_{\text{P}} = 0$ when the dust sublimates.
Also, when there is negligible dust heating from gas collisions,
the cosmic microwave background (CMB)
provides a temperature floor. We include this effect by modifying the
dust radiation rate to $\Lambda_{\text{d}}(T_{\text{d}}) -
\Lambda_{\text{d}}(T_{\text{CMB}})$ (e.g., \citealt{Schneider10}). We
assume that $T_{\text{CMB}} = 50\,$K, corresponding to $z \sim 15$.

\begin{figure} 
\begin{center}
\includegraphics[width=9cm]{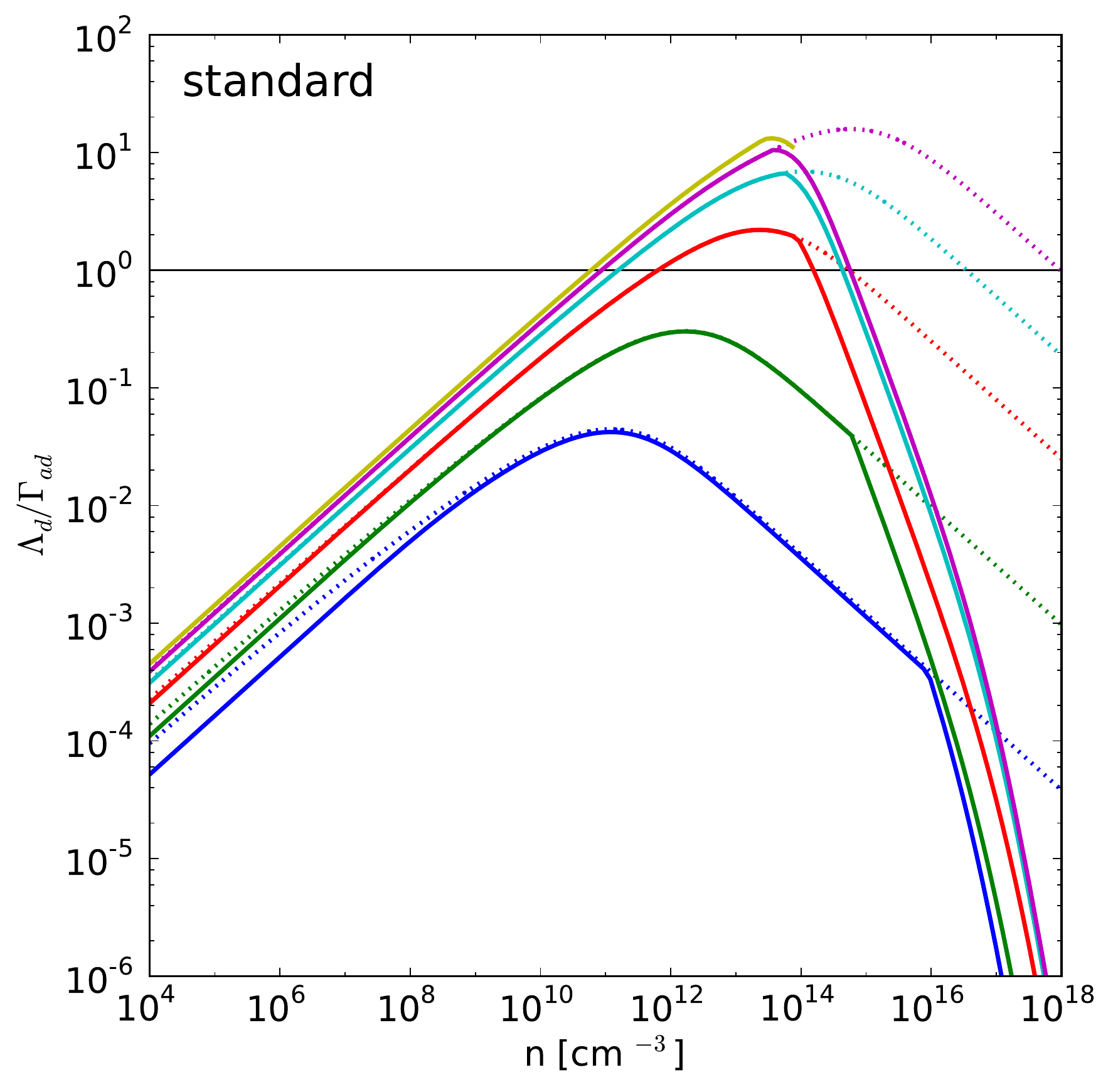} \\
\includegraphics[width=9cm]{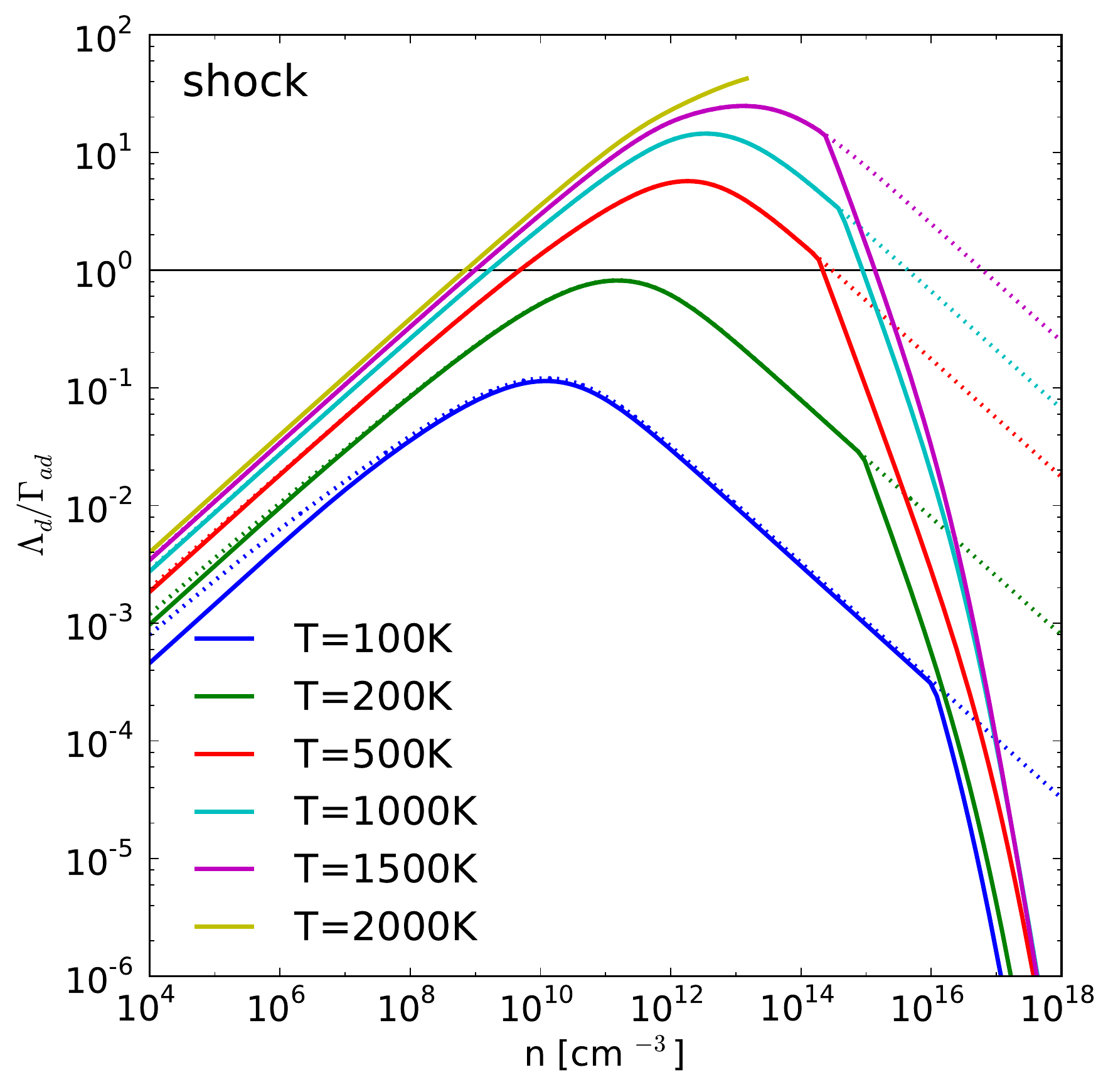}
\end{center}
\caption{\small Ratio between dust cooling rate and adiabatic heating rate as
  a function of gas density and temperature for dust model 1 and
  $\mathcal{D}=10^{-7}$. Top panel shows the 
  standard size distribution, bottom panel the shock size
  distribution. Dotted lines correspond to dust cooling when
  $\beta_{\text{esc}}=1$ while solid lines correspond to dust cooling
  with dust and gas opacity included.
  For $T=2000\,$K, the cooling terminates around
  $n=10^{13}\,$cm$^{-3}$ because the dust sublimates.
  The solid black line indicates where dust cooling is equal to
  adiabatic heating, and the intersection with the colored lines
  indicate densities and temperatures where
  $\mathcal{D}_{\text{crit}}=10^{-7}$.
  \label{fig:opticalthickthin}}
\end{figure}

In summary, the inputs into this model are the gas properties $n$
and $T$; and the dust properties $\kappa_{\text{P}}$, $S$, and
$\mathcal{D}$. The output is a dust temperature $T_{\text{d}}$ with a
corresponding cooling rate $\Lambda_{\text{d}}$.
In Figure~\ref{fig:opticalthickthin}, we show a representative calculation
of $\Lambda_{\text{d}}$ using dust model 1 and
$\mathcal{D}=10^{-7}$. Our simple thermal model is sufficient to
capture many important features of a full thermal evolution 
calculation \citep{Omukai05,Schneider06}. For example, we see that
dust cooling becomes comparable to adiabatic heating at densities
$\gtrsim~10^{10-12}\,$cm$^{-3}$; the smaller grains in the shock size
distribution increase gas cooling; and opacity begins to shut off
dust cooling at densities $\gtrsim~10^{14}\,$cm$^{-3}$.
Note that the $T=2000\,$K lines terminate at 
$n=10^{13}\,$cm$^{-3}$ because the dust sublimates when it reaches
$1500\,$K.

\subsection{Critical Dust-to-Gas Ratio} \label{section:dcrit}
Following \citet{Schneider12}, we define the critical dust-to-gas
ratio $\mathcal{D}_{\text{crit}}$ as the minimum mass fraction of dust
that causes gas fragmentation. We solve for this in a manner similar
to \citet{Bromm03}, by finding the dust to gas ratio such that
\begin{equation} \label{eq:solveforcrit}
  \Lambda_{\text{d}} = \Gamma_{\text{ad}}
\end{equation}
where $\Lambda_{\text{d}}$ is given by Equation \ref{eq:energybalance}
and $\Gamma_{\text{ad}}$ is the adiabatic compressional heating rate,
given by 
\begin{equation} \label{eq:adiabaticheating}
  \Gamma_{\text{ad}} \simeq 1.5 n \frac{k_{\text{B}}T}{t_{\text{ff}}}
\end{equation}
where $t_{\text{ff}}$ is
the free fall time. \citet{Schneider12} show that this method of
finding the dust-to-gas ratio gives a $\mathcal{D}_{\text{crit}}$ that
is very close to a full calculation that accounts for other thermal
effects in the gas.
Also note that the value of $\mathcal{D}_{\text{crit}}$ depends on
the gas density and temperature. 
Following \citet{Schneider12}, we use
a gas density of $n=10^{12}$ cm$^{-3}$ and gas temperature of
$T=1000\,$K as our fiducial values (but see
Section~\ref{section:environments}).

\subsection{Critical Silicon Abundance} \label{section:sicrit}
Given a dust composition with a corresponding $\mathcal{D}_{\text{crit}}$,
we can calculate the minimum amount of silicon required for gas
fragmentation. To do this, we write two expressions for the mass
fraction of Si at the critical point. 

The fraction of silicon in the dust is given by 
\begin{equation}\label{eq:Dcritdust}
  \frac{M_{\text{Si}}}{M_{\text{dust}}}\mathcal{D}_{\text{crit}}
\end{equation}
where $M_{\text{Si}}$ is the mass of silicon in the dust,
$M_{\text{dust}}$ is the total mass of dust, and
$\mathcal{D_{\text{crit}}}$ is the critical dust-to-gas ratio. Note
that $M_{\text{Si}}$ and $M_{\text{dust}}$ depend on the specific dust
model used, and the ratios $M_{\text{Si}}/M_{\text{dust}}$ for our
dust models are given in Table~\ref{tbl:1}. The fraction of silicon in
the gas is given by 
\begin{equation} \label{eq:Dcritgas}
   \frac{\mu_{\text{Si}} \, n_{\text{Si,crit}}}{\mu \, n_{\text{H}}}
\end{equation}
where $\mu_{\text{Si}}$ is the molecular weight of silicon ($28.1 m_{\text{p}}$),
$\mu$ is the molecular weight of the gas, $n_{\text{Si,crit}}$ is the
number density of silicon at the critical point, and $n_{\text{H}}$ is
the hydrogen number density.

We now assume that these two fractions are equal. In other words, we
assume that all silicon present in the gas cloud is locked up in
dust. This maximizes the amount of dust and provides the most
conservative way to calculate a critical silicon threshold.
Setting Equations \ref{eq:Dcritdust} and \ref{eq:Dcritgas} equal and
rewriting them in terms of an abundance, we obtain
\begin{equation} \label{eq:SiHcrit}
  \log \frac{n_{\text{Si,crit}}}{n_{\text{H}}} = 
  \log \mathcal{D}_{\text{crit}} + 
  \log\left(\frac{\mu}{\mu_{\text{Si}}}\right) +
  \log\left(\frac{M_{\text{Si}}}{M_{\text{dust}}}\right)
\end{equation}
and we can find $\mbox{[Si/H]}_{\text{crit}}$ by subtracting the solar
abundances from \citet{Asplund09}\footnote{$\mbox{[X/Y]} =
  \log_{10}(N_X/N_Y)_* - \log_{10}(N_X/N_Y)_\odot$ for element X,Y}. A
star whose measured $\mbox{[Si/H]}$ is less than
$\mbox{[Si/H]}_{\text{crit}}$ thus has a sub-critical $\mathcal{D}$,
too low to trigger dust-induced gas fragmentation.

\begin{figure}
\begin{center}
\includegraphics[width=9cm]{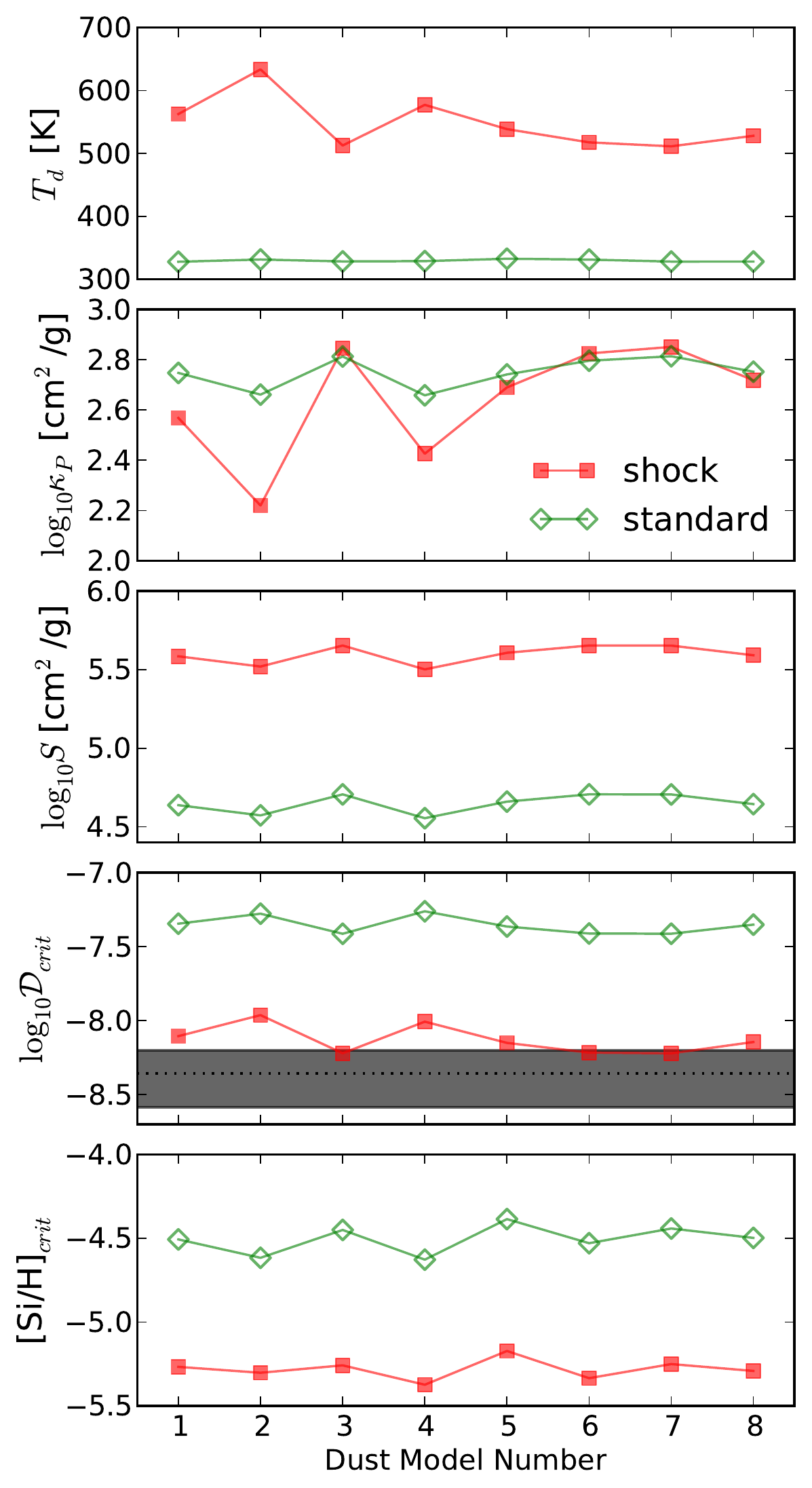}
\end{center}
\caption{Differences in dust properties at the critical point for 
  a gas density of $n=10^{12}$ cm$^{-3}$ and temperature $T=1000\,$K
  for the eight dust models in Table~\ref{tbl:1}.
  From top to bottom: equilibrium dust temperature, Planck mean
  opacity at the equilibrium dust temperature, dust geometric cross
  section, critical dust-to-gas ratio, and critical silicon
  abundance.
  Differences across chemical compositions are relatively
  small, but differences across different size distributions are very
  large.
  The $\mathcal{D}_{\text{crit}}$ from the shock size distribution is
  similar to the $\mathcal{D}_{\text{crit}}$ range from \citet{Schneider12}
  (dotted line and shaded box in fourth panel).
  \label{fig:sizedistr}
}
\end{figure}

In Figure~\ref{fig:sizedistr}, we show the effect of varying the size
distribution and the dust composition on the dust cooling solution at
the fiducial gas density and temperature of $n=10^{12}$ cm$^{-3}$ and
$T=1000\,$K. Table~\ref{tbl:1} shows the numerical values
for $\mbox{[Si/H]}_{\rm crit}$.
The differences between chemical compositions are quite
small, but changing the size distribution makes a very large
difference. In particular, $\mathcal{D}_{\text{crit}}$ and
$\mbox{[Si/H]}_{\text{crit}}$ for the shocked size distribution are
about an order of magnitude lower for all the different chemical
models.
This is a direct result of differences in the average cross section
$S$, as a larger $S$ causes the grains to heat up more quickly
\citep{Schneider06}. In contrast, changing the chemical composition
mostly affects $\kappa_{\text{P}}$, but the steep temperature
dependence of dust cooling (Equation~\ref{eq:lambdad}) implies that
large changes in $\kappa_{\text{P}}$ can be compensated by relatively
small changes in $T_{\text{d}}$.
For comparison, in Figure~\ref{fig:sizedistr} we show the
$\mathcal{D}_{\rm crit}$ calculated in \citet{Schneider12}, where the
dotted line indicates $\mathcal{D}_{\text{crit}} = 4.4 \times 10^{-9}$
and the shaded box indicates $\mathcal{D}_{\rm crit} \in
[2.6,6.3]\times 10^{-9}$. The range in $\mathcal{D}_{\rm crit}$
corresponds to differences just in the grain size distribution/cross
section.
Most of the dust models in \citet{Schneider12} are composed primarily
of carbon dust, and the similarity in $\mathcal{D}_{\rm crit}$ between
these models and our silicate dust models emphasizes that changing the
dust composition produces only a small effect compared to changing the
grain size distribution. 
Many previous authors have also noted the importance of the dust grain
size distribution in determining the cooling properties of dust
(e.g., \citealt{Omukai05,Hirashita09}).

\subsection{Population II Star Forming Environments} \label{section:environments}
The critical dust-to-gas ratio, $\mathcal{D}_{\text{crit}}$, is a function of
the ambient gas density and temperature in the regions where
second-generation, Population~II, star formation takes place.
Their physical conditions are still rather uncertain,
as opposed to the well-defined initial conditions for Population~III
star formation \citep{Bromm13}. 
Thus far we have assumed a fiducial density and temperature of $n
\sim 10^{12}\,$cm$^{-3}$ and $T \sim 1000\,$K where dust cooling will
certainly be important \citep{Omukai05,Schneider12}. This naturally
corresponds to the protostellar disks explored in simulations
\citep{Clark08,Stacy10,Greif11,Clark11,Dopcke13}. However,
other Population~II star forming environments may also
achieve high densities, with likely environments including the
turbulent cores of atomic cooling halos
\citep{Wise07,Greif08,SafShrad12} or in the post-shock region of
a supernova \citep{Mackey03,Salvaterra04,Nagakura09,Chiaki13b}. To
provide a broader view, we consider how dust could impact these
environments by estimating their maximum densities and temperatures.

We do not expect Population~II stars to form in the first
dark matter minihalos since Population~III supernova evacuate much of
the gas from the minihalo, preventing future star formation
\citep{Whalen08}.
However, a $\sim 10^8 M_\odot$ dark matter halo can cool efficiently through
Lyman-$\alpha$ lines \citep{Wise07,Greif08}. These atomic cooling
halos are supersonically turbulent, which can cause densities as high
as $10^6\,$cm$^{-3}$ \citep{SafShrad12}. The virial temperatures of
these halos are quite high ($\sim 10^4\,$K) but H$_2$ cooling can
reduce the temperature to $\sim 400\,$K \citep{Oh02,SafShrad12}. These
conditions will not be sufficient for dust fragmentation
\citep{Omukai05,Schneider12}. However, at the center of these halos
gas can continue collapsing, eventually forming into protostellar disks.

An additional way to obtain a density enhancement is through a
supernova shockwave. The supernova shell and post-shock region can
achieve density enhancements of $10^4$ above the ambient ISM density
\citep{Mackey03}. Thus the maximum density achievable in a shell may
be around $10^6$ cm$^{-3}$, which will again be too low to immediately
fragment through dust cooling. However, shell instabilities may still
cause fragmentation, and subsequent collapse may cause dust-induced
low-mass star formation \citep{Salvaterra04,Nagakura09,Chiaki13b}.

\begin{figure}
\begin{center}
\includegraphics[width=9cm]{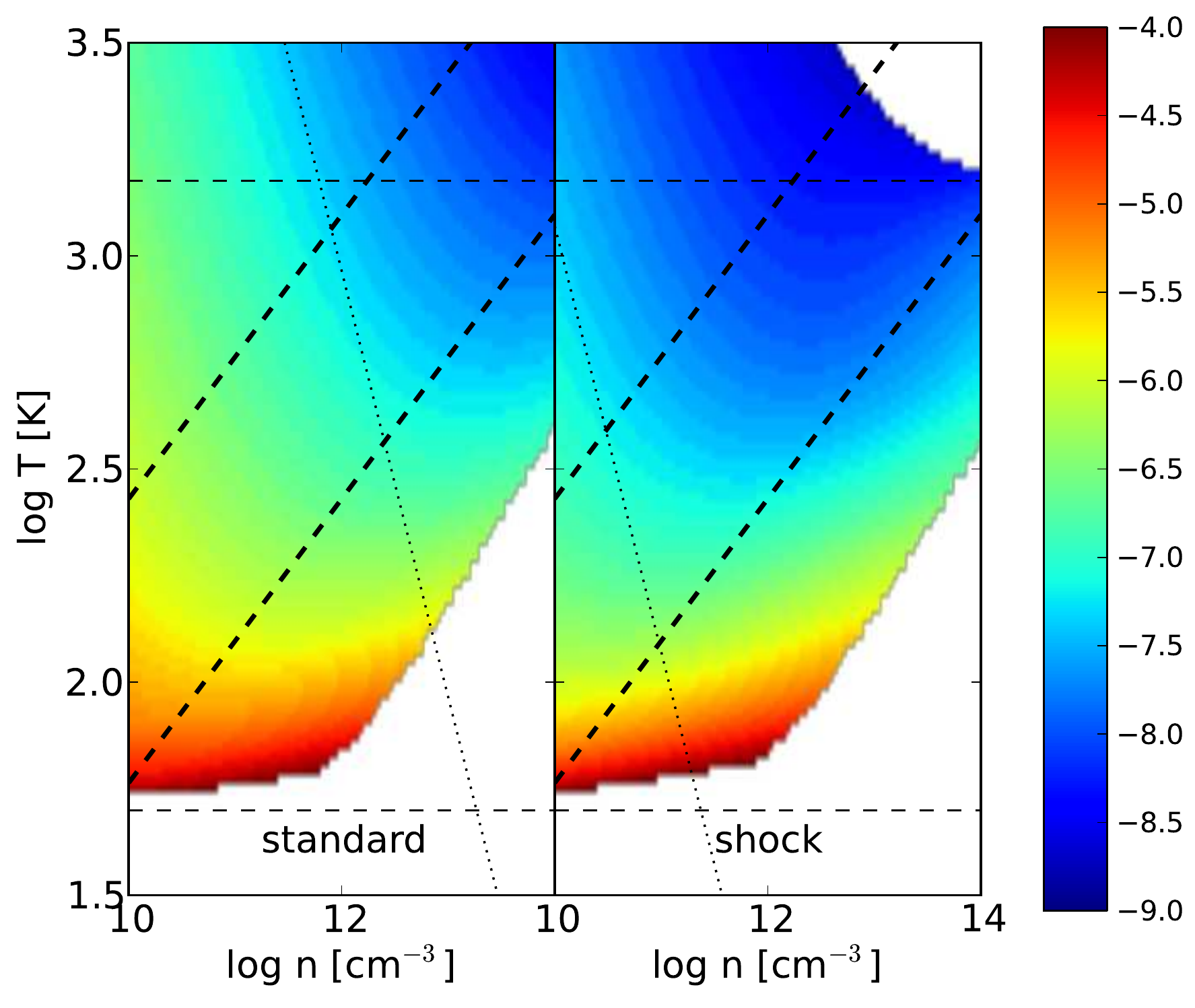}
\end{center}
\caption{$\mathcal{D}_{\text{crit}}$ as a function of gas density and
  temperature for dust model 1. Other dust models are 
  qualitatively similar.
  Left: standard distribution. Right: shock distribution. 
  Horizontal dashed lines indicate the CMB and dust sublimation
  temperatures. 
  Dotted line shows the analytical approximation for
  dust cooling at $\mathcal{D}_{\text{crit}}=10^{-7.5}$ from \citet{Schneider12}.
  Thick slanted dashed lines indicate Jeans masses of $10M_\odot$ and
  $1M_\odot$.
  \label{fig:ntplane}}
\end{figure}

It is clear that in these environments, dust cannot cause widespread
fragmentation until the disk stage of collapse. However, our density
estimates of these environments are rather crude, and future studies
may find other Population~II star forming environments with extremely
high densities. Also, there will certainly be variations in the
density and temperature in a protostellar disk.
Thus, for completeness, we show how $\mathcal{D}_{\text{crit}}$
varies with density and temperature in Figure~\ref{fig:ntplane}.
It is clear that $\mathcal{D}_{\text{crit}}$ (and thus
$\mbox{[Si/H]}_{\text{crit}}$) is somewhat sensitive to the choice of
density and temperature. 
We also show the analytic scaling
of $\mathcal{D}_{\text{crit}}$ derived in \citet{Schneider12} as
dotted lines in Figure~\ref{fig:ntplane} (using $\log
\mathcal{D}_{\text{crit}}= -7.5$). This scaling matches our 
calculation well at higher gas temperatures, as expected based on the
approximation $T_{\text{d}}=0$ used to derive the formula.

We note that we use a simple thermal model that only considers
adiabatic heating and dust thermal cooling. Thus, the
$\mathcal{D}_{\text{crit}}$ values in Figure~\ref{fig:ntplane} should
be treated as guidelines that approximate what would be obtained from
a more complete thermal model or from simulations.

\section{Comparison with Metal-Poor Star Abundances} \label{section:data}
We now compare the critical silicon abundances for star forming gas,
as derived from all eight of our silicon-based dust models with both
grain size distributions, to abundance measurements of metal-poor
stars.
We evaluate $\mbox{[Si/H]}_{\text{crit}}$ at $n=10^{12}\,$cm$^{-3}$
and $T=1000\,$K (Figure~\ref{fig:sizedistr}, Table~\ref{tbl:1}).

\begin{figure}
\begin{center}
\includegraphics[width=9cm]{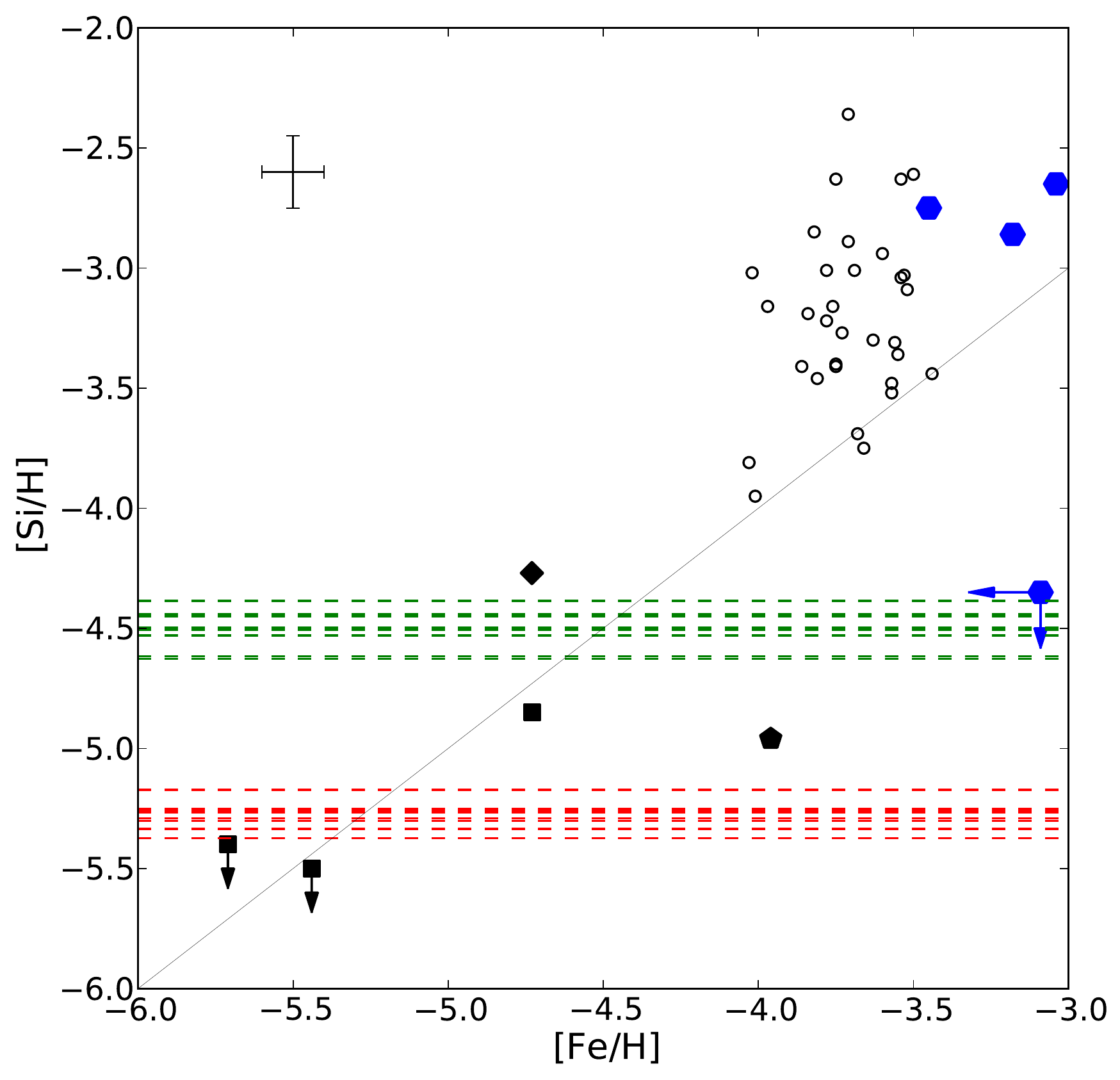}
\end{center}
\caption{Silicon and iron abundances for our sample compiled from the
  SAGA database \citep{Suda08,Frebel10,Yong13}.
  We include our new silicon abundance measurement for
  HE~0557$-$4840 and upper limits for HE~0107$-$5240 and
  HE~1327$-$2326. We show typical errors on the abundance measurements
  in the top left corner. We plot critical silicon abundances
  calculated for $n=10^{12}$ cm$^{-3}$ and $T=1000\,$K, indicated by
  the dashed horizontal lines. The green dashed lines are computed
  using the standard size distribution, and the red lines are computed
  with the shock size distribution.  The black line indicates
  $\mbox{[Si/Fe]}=0$ as a reference.  Five stars are emphasized by
  larger black symbols.  The black squares are, from low to high
  $\mbox{[Fe/H]}$: HE~1327$-$2326 \citep{Frebel08}, HE~0107$-$5240
  \citep{Christlieb04}, and HE~0557$-$4840 \citep{Norris07}. The black
  pentagon is HE~1424$-$0241 \citep{Cohen08}.  The black diamond is
  SDSS~J1029151+1729 \citep{Caffau11}. The blue hexagons show the
  three most iron poor DLAs from \citet{Cooke11} and the upper limits
  from \citet{Simcoe12} (see Section~\ref{section:DLA}).
  \label{fig:sih} }
\end{figure}

We use metal poor halo stars and dwarf galaxy stars with
$\mbox{[Fe/H]} < -3.5$ taken from the literature
\citep{Suda08,Frebel10,Yong13}. References to individual abundances
can be found for all but the most iron-poor stars in the SAGA database
\citep{Suda08}.
Figure~\ref{fig:sih} shows $\mbox{[Si/H]}$ as a
function of $\mbox{[Fe/H]}$ for our stars.  For consistency, we use
abundances derived from 1D LTE stellar atmosphere models (but see
further discussion below). The horizontal dashed lines indicate the
critical silicon abundances from our dust models. The lines are
colored by size distribution: green lines correspond to the
standard size distribution, and red lines correspond to the shock
size distribution. As previewed in Figure~\ref{fig:sizedistr}, the
critical silicon abundances are higher for the standard size
distribution by almost an order of magnitude, but variation between
different chemical compositions is relatively low and less than
$0.3\,$dex. For reference, we also show the solar silicon-to-iron
ratio as a thin black line.  As can be seen, the stellar abundances
cover a large range in the diagram. Stars with $\mbox{[Fe/H]} > -4.0$
have typical $\alpha$-abundance ratios of $\mbox{[Si/Fe]} \sim 0.4$
and higher, albeit with one exception.  For this study, stars with
$\mbox{[Fe/H]} \lesssim -4.5$ or $\mbox{[Si/H]} \lesssim -4.5$ are of
particular interest. Indeed, there are several objects in this range which
we use as test objects for our modeling of dust cooling in the earliest star
forming environments. The higher metallicity stars are unfortunately
not usable in this context as they likely formed at a later time 
from gas that already contained enough metals for cooling.

We note that silicon abundance measurements can be challenging in the
most metal-poor stars given the overall weakness of absorption lines.
Moreover, the strongest Si line at $3905\,${\AA} is blended with a
molecular CH line. As most of these stars are carbon-enhanced, Si
abundances or upper limits are difficult to derive.  As a result,
HE~0557$-$4840 ($\mbox{[Fe/H]}=-4.7$, \citealt{Norris07,Norris12}),
HE~1327$-$2326 ($\mbox{[Fe/H]}=-5.7$,
\citealt{Frebel05,Frebel06,Frebel08}), and HE~0107$-$5240
($\mbox{[Fe/H]}=-5.4$,
\citealt{Christlieb02,Christlieb04,Bessell04}) do not have published
silicon abundances or (tight) upper limits. From having available
spectra of these objects, we used the spectrum synthesis technique
(see e.g., \citealt{Frebel13a} for further details) and published
stellar parameters and carbon abundances
\citep{Norris07,Frebel05,Christlieb02} to derive a silicon abundance
for HE~0557$-$4840 and upper limits for HE~1327$-$2326 and
HE~0107$-$5240.  For HE~0557$-$4840, the silicon line is somewhat
distorted in addition to the carbon blend, but two different spectra
yield a consistent result of $\mbox{[Si/H]} = -4.85 \pm 0.2$.  For
HE~0107$-$5240 and HE~1327$-$2326, a visual examination of the spectra
shows no apparent absorption at $3905\,${\AA}, although again there is
a strong CH feature very close to the position of the Si line. Our
newly determined upper limits are $\mbox{[Si/H]} < -5.5$ for
HE~0107$-$5240 and $\mbox{[Si/H]} < -5.4$ for HE~1327$-$2326. In the
case of HE~0107$-$5240, we thus found a much improved limit compared to
an equivalent width-based upper limit \citet{Christlieb04}.

Before comparing our critical silicon abundances to those observed in
the metal-poor stars, it is important to briefly consider effects on
abundances derived from 1D LTE model atmospheres, which can yield
different abundances compared to using more physical 3D LTE
hydrodynamic models or carrying out additional NLTE
corrections. Although the carbon and oxygen abundances derived from 3D
models have abundance corrections of order $+0.5\,$dex, the available
3D iron and silicon abundances appear to be within $+0.2\,$dex of the
1D abundances \citep{Collet06,Caffau11}.  However, NLTE effects on 1D
abundances can increase the silicon abundances in metal-poor stars by
$0.2$ to $0.5\,$dex when $T_{\text{eff}} > 5500\,$K. This effect
becomes larger as stars become hotter \citep{Shi09,Zhang11}. Of the
interesting stars, HE~1327$-$2326 ($T_{\text{eff}}=6180\,$K) and
SDSS~J1029151+1729 ($T_{\text{eff}}=5811\,$K) may be affected. But
since only the most iron-poor stars have 3D LTE abundances available,
we show the 1D LTE abundances of all stars in Figure~\ref{fig:sih}. We
then assume that within the given error bars, these abundances are
reasonably accurately describing the Si and Fe content of the stars,
especially relative to each other.

In Figure~\ref{fig:sih}, the three black squares are HE~1327$-$2326,
HE~0107$-$5240, and HE~0557$-$4840. These stars all have silicon
abundances that fall below the critical lines for the standard size
distribution, showing that they could not have formed from gas cooled
by silicon-based dust of this size distribution. Furthermore, HE~0107$-$5240 and
HE~1327$-$2326 have silicon upper limits that are even slightly below
the critical silicon abundances derived from the shock size
distribution. 
This suggests that both of these stars did not form
because of the agency of silicon-based dust cooling at all, but
instead relied on some other mechanism to enable low-mass star
formation.

The star HE~1424$-$0241 is also interesting because of an abnormally
low silicon abundance, $\mbox{[Si/Fe]}=-1.00$, despite its somewhat
higher iron abundance ($\mbox{[Fe/H]}=-3.96$; \citealt{Cohen08})
compared to the stars described above. It also has only an upper limit
on the carbon abundance and anomalously low $\mbox{[Ca/Fe]}$ and
$\mbox{[Ti/Fe]}$ abundance, but significant enhancements in
$\mbox{[Mn/Fe]}$ and $\mbox{[Co/Fe]}$. This star is shown as the black
pentagon in Figure \ref{fig:sih}. It also falls beneath the critical
silicon abundances derived from the standard size distribution. While
this is certainly interesting in the context of testing for cooling
mechanisms, it may be possible that this star's abundance
pattern does not reflect nucleosynthesis products of typical
supernovae, as such a low Si abundance has never before been found in
similar metal-poor stars \citep{Cohen08}.

There is another interesting star with low iron, SDSS~J1029151+1729
(\citealt{Caffau11}, $\mbox{[Fe/H]}=-4.73$, black diamond in
Figure~\ref{fig:sih}). It has $\mbox{[Si/H]} = -4.3$ which places it
above the critical silicon values for all of our models. This star is
also not carbon-enhanced (see further discussion in
Section~\ref{section:dustvsFS}) and it has previously been suggested
that this star formed from dust-cooled gas
\citep{Schneider12b,Klessen12}. Our results agree with this finding.

Overall, from Figure~\ref{fig:sih}, it is apparent that within our
framework, the four stars falling beneath the standard size
distribution's critical silicon abundances are unable to have formed in
a cloud cooled by silicate dust with a Milky Way grain size distribution. Thus,
the fragmentation seen in simulations using metallicity-scaled Milky Way
dust (e.g., \citealt{Omukai10,Dopcke13}) cannot explain the formation
of these four stars\footnote{As a consistency check: 
  we calculate $\mathcal{D}_{\text{crit}}$ of $\sim 10^{-7.5}$ for the
  standard size distribution. The Milky Way dust-to-gas ratio is $\sim
  10^{-2}$ \citep{DraineBook}. Thus, when scaling by $Z/Z_\odot =
  10^{-5}$ this is above $\mathcal{D}_{\text{crit}}$, but when scaling
  by $Z/Z_\odot =  10^{-6}$ this is below
  $\mathcal{D}_{\text{crit}}$. This matches the simulation results
  of \citet{Omukai10} and \citet{Dopcke13}.}.
It follows that either this type of dust is not an accurate model of
dust in the early universe, or that the presence of such dust in early
gas clouds was subject to stochastic events (e.g., individual
supernovae), only rarely leading to the cooling required for star
formation to occur.

\section{Two pathways for early low-mass star formation?} \label{section:dustvsFS}
In Section~\ref{section:data}, and assuming the suppression of
carbon-based dust, we found that some stars apparently cannot form
from gas cooled by only silicon-based dust.
In a broader context, it is then interesting to consider the relative
importance of dust thermal cooling and carbon/oxygen fine structure
line cooling. We can derive new constraints on a star's formation
process by considering its silicon abundance in conjunction with
$D_{\text{trans}}$ from \citet{Frebel07}. Hence, we calculate
$D_{\text{trans}}$ for our star sample with the updated formula from
\citet{Frebel13a}:
\begin{equation} \label{eq:dtrans}
  D_{\text{trans}} = \log(10^{\mbox{[C/H]}}+0.9 \times 10^{\mbox{[O/H]}})
\end{equation}
To emphasize our notation, note the difference between $\mathcal{D}$ which
represents a dust-to-gas ratio, and $D_{\text{trans}}$ which is the transition
discriminant of \citet{Frebel07}. 

\begin{figure}
\begin{center}
\includegraphics[width=9cm]{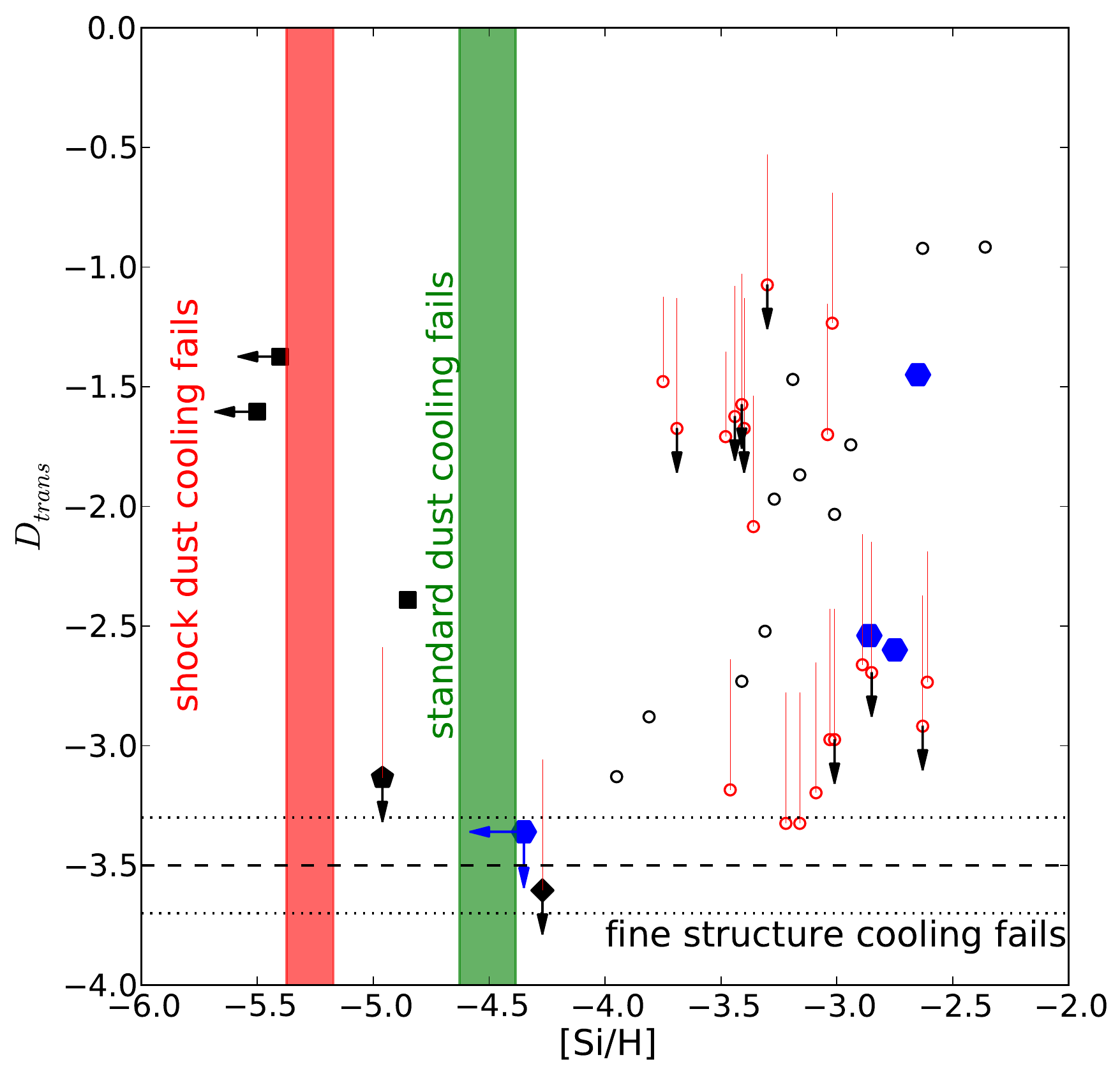}
\end{center}
\caption{$D_{\text{trans}}$ vs $\mbox{[Si/H]}$ for our sample of stars
  using 1D LTE abundances.
  The shaded red and green bars are the range of critical silicon
  abundances shown in Figure \ref{fig:sih}. The critical $D_{\text{trans}}$
  value and errors are shown as dashed and dotted horizontal black
  lines.
  The four stars that fall to the left of the green bar likely cannot
  form through dust cooling, while the star that falls below the
  dashed line likely cannot form through fine structure cooling.
  This is evidence that both dust cooling and fine structure cooling
  can be relevant for low mass star formation. It is also tentative
  evidence that fine structure cooling and dust cooling are mutually
  exclusive.
  \label{fig:dtranssih}}
\end{figure}

In Figure~\ref{fig:dtranssih}, we show $D_{\text{trans}}$ as a
function of the silicon abundance. Stars that have both carbon and
oxygen abundances available are plotted in black. Following
\citet{Frebel13a}, stars missing either carbon or oxygen are plotted
in red, with a vertical bar denoting the $D_{\text{trans}}$ range
corresponding to $-0.7<$[C/O]$<+0.2$.

The four stars with the lowest silicon abundances appear to all have
large carbon abundances, placing them above the critical
$D_{\text{trans}}$ value of $-3.5$. SDSS~J1029151+1729 however has a
relatively high silicon abundance (at $\mbox{[Si/H]} = -4.3$) and a
low carbon abundance placing it below the critical $D_{\text{trans}}
=-3.5$ level. This combination of low Si/high C and high Si/low C
abundances is an interesting finding which warrants further exploration
in future work. However, if dust in the early universe is
silicon-based, then the currently available data suggest a bifurcation
in the dominant cooling mechanisms of the gas clouds that produced
these low-mass stars.
This lends support to the arguments made by \citet{Norris13} who
suggest different paths of star formation for carbon-enhanced
metal-poor stars ($\mbox{[C/Fe]} >0.7$) and those that do not show
such a significant overabundance of carbon relative to iron. 
After all, nearly a quarter of extremely metal-poor stars with
$\mbox{[Fe/H]} < -2.5$ are carbon-enhanced (e.g., \citealt{Beers05}),
with the carbon-rich fraction increasing with decreasing [Fe/H].

\begin{figure}
\begin{center}
\includegraphics[width=9cm]{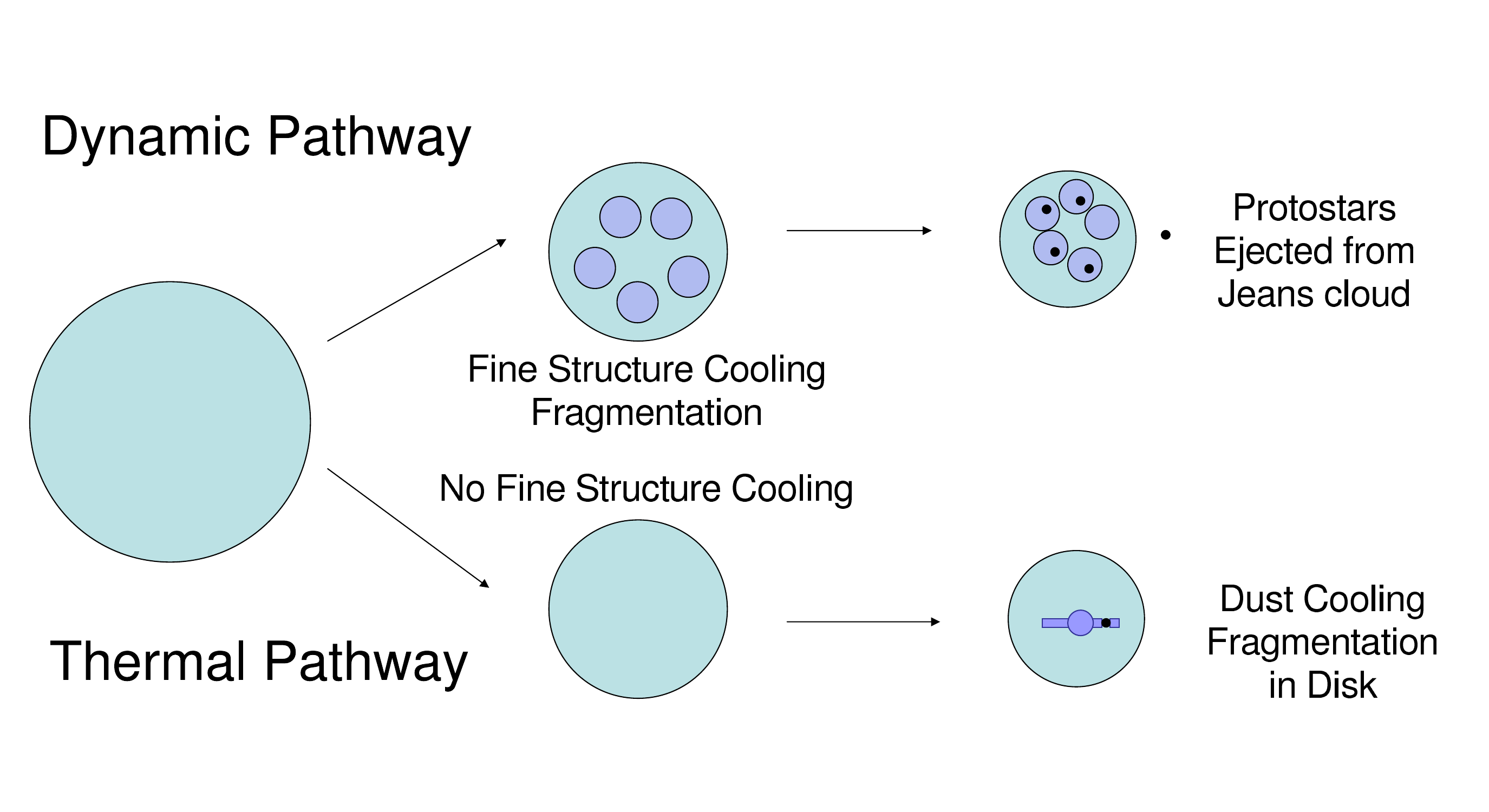}
\end{center}
\caption{
  Two potential pathways for low-mass metal-poor star formation. We
  start with a collapsing gas cloud on the left.
  In the dynamic pathway, fine structure cooling induces vigorous
  fragmentation into many sub-clumps (e.g.,
  \citealt{Bromm01}). Many-body dynamics can then cause the ejection
  of a protostar from its parent cloud, creating a low-mass star
  without dust cooling. 
  In the thermal pathway, the absence of fine
  structure cooling causes the entire cloud to collapse without
  experiencing subfragmentation.
  The center of the cloud forms a protostellar disk with
  high density, and dust cooling causes low-mass fragmentation in the
  disk (e.g., \citealt{Dopcke13}).
  \label{fig:cartoon}}
\end{figure}

We thus further examine the physics driving these two potential
pathways, which may guide future work towards clarifying emerging
bimodal picture of first low-mass star formation.
We start by considering gas collapse within an atomic cooling halo
\citep{Wise07,Greif08}. Here, a bifurcation occurs into two different
pathways depending on the fragmentation properties of the gas. A
schematic view of these two pathways is shown in Figure
\ref{fig:cartoon}.

In the first pathway, a large collapsing gas cloud undergoes vigorous
fragmentation into many medium-mass clumps \citep{Bromm01,SafShrad13}.
The presence of a LW background from the first stars (e.g.,
\citealt{Ciardi00}) prevents molecular hydrogen from dominating the
cooling rate, and thus fine structure cooling is required to enable
fragmentation at these intermediate densities \citep{SafShrad13}.
The result is a strongly clustered star formation mode, where typical
stars may grow to masses $\gtrsim 10M_\odot$, but also leading to a
retinue of lower-mass cluster members. Specifically,
many-body gravitational interactions may eject some of these
protostars from their parent clouds, thus shutting-off further accretion,
so that they remain at low masses.
We call this mode the ``dynamic pathway'', which could be reflected in the
carbon-enhanced metal-poor stars.
Alternatively, the atomic carbon may condense into dust grains
at high densities, inducing gas fragmentation \citep{Chiaki13a}.

The second pathway involves monolithic collapse of a Jeans-unstable
gas cloud. In the absence of significant fine-structure cooling, the
gas just continues to collapse until a protostellar disk forms at
the center of the cloud (e.g., \citealt{Clark08}). The LW
background prevents fragmentation at intermediate densities from
molecular hydrogen cooling \citep{SafShrad13}.
In the disk, the density is high enough for dust cooling to be
significant, and the disk fragments into low-mass clumps
\citep{Dopcke13}. We call this the ``thermal pathway'' and
note that rotation support
is critical for providing an environment that is stable for longer
than the gravitational free fall time \citep{Tohline80,Clark08}.
Although the LW background inhibits fragmentation from molecular
cooling, it is possible that other processes could cause additional
fragmentation away from the center of the halo. For example, a
shell instability in a supernova shock may cause fragmentation,
creating additional star clusters in the atomic cooling halo
(e.g., \citealt{Salvaterra04,Nagakura09,Chiaki13b}).

While our two-pathway interpretation is still largely qualitative at
this stage, the current body of metal-poor stellar abundance data can
only be satisfactorily explained with such different star formation
processes occuring in the early universe. Future modeling of gas
cooling and metal mixing processes will shed more light on the matter.
Additional discoveries of metal-poor stars with iron, carbon, oxygen,
and silicon abundance measurements and upper limits will greatly help
to confirm or refute this two-pathway theory by populating the
parameter space presented in Figure~\ref{fig:dtranssih}.

\section{Damped Lyman-$\alpha$ Hosts} \label{section:DLA}
Chemical abundances of Damped Lyman-$\alpha$ (DLA) systems have
potential to help us understand the star formation 
environment that may have hosted these early metal-poor stars.
DLA's have indeed been hypothesized to be observational probes of the
environment where metal-poor Population II stars may form (e.g.,
\citealt{Cooke11} and references within). They may also be able to
constrain the Population III initial mass function \citep{Kulkarni13}.
Most DLAs observed to date have $\mbox{[Fe/H]} > -3.5$
\citep{Cooke11}, but recently a high-redshift DLA candidate has been
discovered with only upper limits on metal abundances
\citep{Simcoe12}. There has also been evidence that gas may remain
very pristine at lower redshifts as well \citep{Fumagalli11}.

We show the chemical abundances of the three most iron-poor DLAs from
\citet{Cooke11} and the upper limits from \citet{Simcoe12} as blue
hexagons in Figures \ref{fig:sih} and \ref{fig:dtranssih}. The three
DLAs from \citet{Cooke11} have abundances that fall within the scatter
of the more metal-rich stars of our sample, which is consistent with
the interpretation that these DLAs could be the formation sites of the
metal-poor stars in our halo. The DLA candidate from \citet{Simcoe12}
has abundance limits at the critical values of both the fine structure
and dust cooling criteria. This could be interpreted such that neither
dust nor fine-structure line cooling have operated in this system,
leading to no low-mass star formation. However, the
nature of this DLA remains somewhat ambiguous \citep{Simcoe12}, so
this interpretation may need to be revised. Future observations of
metal-poor DLAs will show whether additional systems can be
found with such low abundances, and whether any will be below the
critical silicon and carbon/oxygen abundances as presented in this
paper. In fact, more metal-poor DLA's would greatly help to further
constrain the formation environment of the most metal-poor stars
in the Milky Way halo.

\section{Caveats} \label{section:caveats}
\subsection{Impact of Carbon Dust} \label{section:cdust}
The most important assumption in our work is that dust in the early
universe is largely silicon-based.  If large amounts of non-silicate
dust is produced, then the critical silicon abundance may not be
suitable for testing dust cooling with the most metal-poor stars.
In particular, as mentioned in
Section~\ref{section:whynocarbon} and discussed in \citet{Cherchneff10},
significant amounts of carbon dust may form if carbon-rich regions are
not microscopically mixed with helium ions in the supernova ejecta.
\citet{Cherchneff10} calculate an upper limit on carbon dust
produced in this situation by assuming no mixing between the carbon
and helium layers. 95\% of the carbon-rich/oxygen-poor layer is
depleted for a total of $0.0145 M_\odot$ of carbon dust, or about
10\% of the final dust mass in dust models 1 and 2.

\begin{figure*}
\begin{center}
\includegraphics[width=18cm]{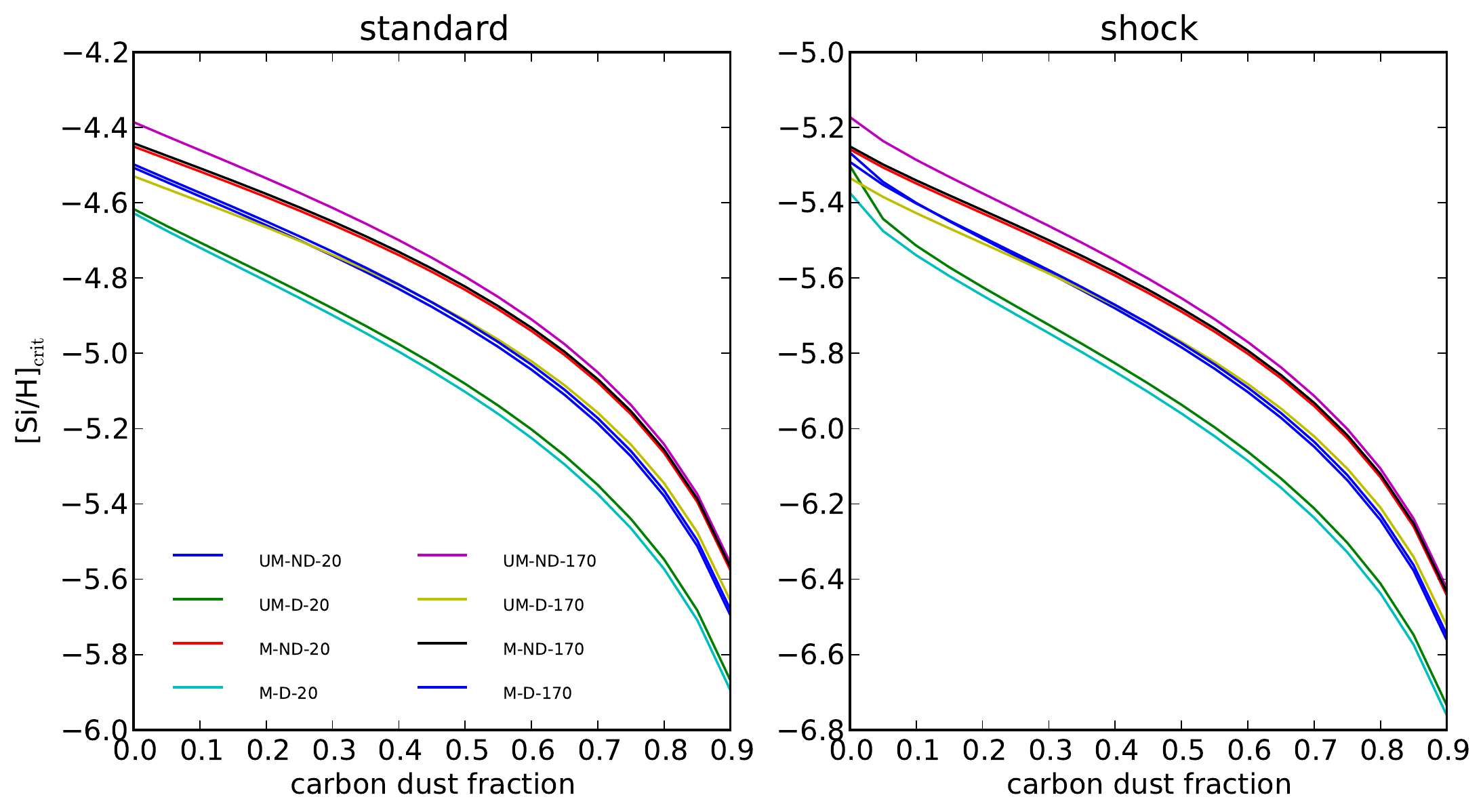}
\end{center}
\caption{
  Critical silicon abundance for our eight dust models as a function
  of carbon dust fraction.
  The general shape is dominated by the silicon mass term in
  Equation~\ref{eq:SiHcrit}, with only minor contributions from the
  change in $\mathcal{D}_{\rm crit}$.
  \label{fig:carbfrac}}
\end{figure*}

The level of mixing, and thus how much carbon dust is
produced, depends on many variables including the nature of the
supernova.  Thus, we recompute $\mbox{[Si/H]}_{\rm crit}$ for our
dust models after adding different amounts of carbon dust directly
to the dust models in Table~\ref{tbl:1}.  The results are shown in
Figure~\ref{fig:carbfrac}.  
The general shape is logarithmic, corresponding to the
silicon mass term in Equation~\ref{eq:SiHcrit}. Changes in
$\mathcal{D}_{\rm crit}$ affect $\mbox{[Si/H]}_{\rm crit}$ mostly at
low carbon fractions.

When $\sim$20\% of the dust mass is in carbon, there is a
$\sim$0.2 dex shift down in $\mbox{[Si/H]}_{\rm crit}$ (see
Figure~\ref{fig:carbfrac}). This does not significantly affect 
our conclusions from Section~\ref{section:data}. However, it
complicates our interpretation in Section~\ref{section:dustvsFS}
(see Figure~\ref{fig:dtranssih}) as the carbon cannot be directly
associated with fine structure cooling.  Above $\sim$50\% dust mass
in carbon, $\mbox{[Si/H]}_{\rm crit}$ shifts down by $\gtrsim$0.5
dex. As a consequence, silicon is no longer a useful element
for empirically evaluating the role of dust.

In principle, the methodology described in
Section~\ref{section:sicrit} could be applied to derive critical
abundances for carbon dust.  An upper limit on the critical carbon
abundance can be found by assuming pure carbon dust. We find
$\mbox{[C/H]}_{\rm crit, max} \sim-4.9$ for the standard size
distribution and $\sim-5.8$ for the shock size distribution.  These
thresholds are so low that falsifying a carbon dust theory is
observationally intractable at the present time.  We estimate that
$\mbox{[C/H]} \lesssim -5$ could be measured for a suitable bright,
cool ($T \sim 4600\,$K) giant if the signal-to-noise is over
300. This is at the edge of current telescope capabilities, but
spectrographs on the next generation of extremely large telescopes
(e.g., GCLEF on GMT) should enable observations of extremely low
carbon abundances. Thus, although testing carbon dust with a
critical carbon criterion is currently impractical, it may be
accessible in the future.

\subsection{Other Considerations}
Unlike \citet{Schneider12b}, we do not fit separate supernovae yields
to individual stellar abundance patterns. However, we have verified
that the supernovae abundances fall within the abundance range of
known metal-poor stars \citep{Frebel10}. Different supernova yields
would affect the dust compositions computed by the
\citet{Cherchneff10} models. In particular, if the ejecta were
to be very abundant in carbon, regions of unmixed carbon are more
probable and thus larger amounts of carbon dust would form.

We chose two simple grain size distributions to derive our critical
silicon abundances instead of calculating them specifically for our
dust models.
In doing so, we made the simplifying assumption
that all types of dust follow the same size distribution. This
assumption is likely not accurate as different chemical species
condense to different initial sizes and undergo different amounts of
destruction in a supernova reverse shock (e.g.,
\citealt{Todini01,Bianchi07,Nozawa07,Silvia10}).
Since different grain chemical species may not be in thermal
equilibrium with each other due to their low density, it may be
important to treat grain types separately instead of lumping them
together into a single dust model as is typically done in the
literature as well as in this paper.
We also note that although smaller dust grains lead to more efficient gas
cooling, the supernova reverse shocks generally responsible for
breaking up dust grains also completely destroy a significant fraction
of the dust \citep{Bianchi07,Silvia10}.

We did not consider grain growth, which can create significantly more
dust \citep{Chiaki13a}. However, at low metallicities, this is not
important for our fiducial density of $10^{12}\,$cm$^{-3}$ \citep{Hirashita09}.
We also neglected the effect of increased H$_2$ formation on the
surfaces of dust grains. However, the increased H$_2$ cooling should
be roughly balanced by the heat released in forming H$_2$ (e.g.,
\citealt{Omukai10,Glover13,Dopcke13}). We do not expect that including 
this effect would significantly change the critical silicon
abundances, but it may be relevant for causing additional
fragmentation in the thermal pathway \citep{SafShrad13}.

\section{Conclusion} \label{section:conclusion}
We have computed critical silicon abundances using the
  silicon-based dust models in \citet{Cherchneff10}. 
  We found that different dust chemical compositions introduce only
  small variations ($\sim 0.2$ dex) in the critical silicon abundance,
  but assumptions about the size distribution can produce an order of
  magnitude difference, with smaller grains being much more effective
  at cooling the gas (Figure~\ref{fig:sizedistr}).
At the densities and temperatures associated with protostellar
  disks, the critical silicon abundance is
  $\mbox{[Si/H]}=-4.5 \pm 0.1$ for a standard Milky Way grain size
  distribution and $\mbox{[Si/H]}=  -5.3 \pm 0.1$ for a shocked
  grain size distribution.
  Other Population~II star forming environments are not likely to be
  influenced by dust because their densities are too low.

We then compare our critical silicon abundances to chemical
abundances of metal-poor stars.
For the standard Milky Way grain size distribution,
four of the nine stars with $\mbox{[Fe/H]} < -4.0$ and three of
the four stars with $\mbox{[Fe/H]} \lesssim -4.5$ have silicon
abundances too low to be explained by silicon-based dust
cooling. All stars that cannot form through silicon-based dust
cooling satisfy the $D_{\text{trans}}$ criterion, with the
possible exception of HE~1424$-$0241. (Figures~\ref{fig:sih}
and~\ref{fig:dtranssih}).

In fact,
two stars have silicon abundances below even the critical
silicon abundances for the shocked size distribution, suggesting
that silicon-based dust may not have played a dominant
role in their formation.
With the caution required in interpreting a small sample of
stars, we thus see hints of two distinct pathways for the
formation of low-mass metal-poor stars in the early universe.
One pathway depends on fine structure cooling, and the other
depends on dust cooling (Figure~\ref{fig:cartoon}). 

The most important uncertainty in this analysis is the
production of carbon dust, which can occur if carbon-rich regions of
supernova ejecta are not microscopically mixed with helium ions.
If significant amounts of carbon dust can form, the critical
silicon abundance will decrease (Figure~\ref{fig:carbfrac}). If carbon
dust is less than 20\% of the total dust mass, the critical silicon
abundances shift by less than 0.2 dex and our comparison with data is
not significantly affected. However, if more of the dust is in carbon,
the critical silicon abundance may not be a good criterion to evaluate
dust cooling and our interpretation of Figures~\ref{fig:sih}
and~\ref{fig:dtranssih} may need revisiting.
A more complete understanding of microscopic mixing and dust formation
in Population III supernova ejecta may allow us to better determine a
carbon dust fraction.

Given these results, we note that many potentially interesting
metal-poor stars in the literature do not have silicon abundances measured.
We encourage observers to consider measurements of silicon abundances
or upper limits, both in future data and in currently available
spectra. Additional discoveries of metal-poor DLAs may furthermore
help to understand the birth clouds of metal-poor stars in the early
universe.  Only with more data can we observationally evaluate this
and other potential models for the formation of the most metal-poor
stars.

\section*{Acknowledgements}
We thank an anonymous referee for improving the manuscript,
especially regarding our treatment of the dust models.
We thank Ralf Klessen and Naoki Yoshida for conversations
regarding dust properties,
Chalence Safranek-Shrader and Gen Chiaki for clarifying many
points regarding their work, and
 John Norris and Norbert Christlieb for providing us with
spectra of HE~0557$-$4840 and HE~0107$-$5240.
A.J and A.F. are supported by NSF grant AST-1255160. V.B. is supported by NSF
grant AST-1009928 and by NASA ATFP grant NNX09-AJ33G.

\clearpage
\begin{turnpage}
\begin{deluxetable}{ccrrrrrrrrrrrr}
\tabletypesize{\tiny}
\tablecolumns{14}
\tablecaption{Dust Model Masses from \citet{Cherchneff10}}
\tablewidth{0pt}
\tablehead{
  \colhead{ID} & \colhead{Model Name} &
  \colhead{SiO$_2$} & 
  \colhead{Fe$_2$SiO$_4$} & 
  \colhead{Mg$_2$SiO$_4$} & 
  \colhead{Si} &
  \colhead{Fe} & 
  \colhead{FeS} & 
  \colhead{Mg} & 
  \colhead{MgO} &
  \colhead{Al$_2$O$_3$} & 
  \colhead{M$_{\text{Si}}$/M$_{\text{dust}}$} &
  \colhead{$\mbox{[Si/H]}_{\rm crit, standard}$} &
  \colhead{$\mbox{[Si/H]}_{\rm crit, shock}$}
  \label{tbl:1}
}
\startdata
1 & UM ND 20\tablenotemark{a} & 0.039 & 0 & 0 & 0.030 & 4.6 $\times 10^{-5}$ & 0.033 &
3.9 $\times 10^{-4}$ & 3.9 $\times 10^{-4}$ & 8.6 $\times 10^{-5}$ &
0.469 & -4.51 & -5.27 \\
2 & UM D 20\tablenotemark{a}  & 0 & 0 & 0.089 & 0.030 & 4.6 $\times 10^{-5}$ & 0.033 &
3.9 $\times 10^{-4}$ & 3.9 $\times 10^{-4}$ & 8.6 $\times 10^{-5}$ &
0.312 & -4.62 & -5.30 \\
3 & M ND 20  & 0.105 & 0 & 0 & 0.049 & 4.3 $\times 10^{-4}$ & 0 &
1.4 $\times 10^{-3}$ & 0 & 8.8 $\times 10^{-4}$ &
0.625 & -4.45 & -5.26 \\
4 & M D 20   & 0 & 0.125 & 0.160 & 0.049 & 0 & 0 &
0 & 0 & 8.8 $\times 10^{-4}$ &
0.293 & -4.63 & -5.37 \\
5 & UM ND 170\tablenotemark{a} & 3.638 & 0 & 0 & 1.963 & 6.7 $\times 10^{-5}$ & 0.011 &
8.02 $\times 10^{-3}$ & 2.5 $\times 10^{-6}$ & 0.0297 &
0.648 & -4.39 & -5.17 \\
6 & UM D 170\tablenotemark{a}  & 2.577 & 0 & 2.474 & 1.963 & 6.7 $\times 10^{-5}$ & 0.011 &
2.5 $\times 10^{-4}$ & 2.5 $\times 10^{-6}$ & 0.0296 &
0.519 & -4.53 & -5.33 \\
7 & M ND 170  & 17.3 & 0 & 0 & 8.1 & 0.004 & 0 & 0 & 0 & 0.003 & 
0.637 & -4.44 & -5.25 \\ 
8 & M D 170   & 12.9 & 6.6 & 5.7 & 8.1 & 0.004 & 0 & 0 & 0 & 0.003 & 
0.486 & -4.50 & -5.29 \\ 
\enddata
\tablecomments{Dust masses are in $M_\odot$. Model names refer to the
  type of dust model from \citet{Cherchneff10}. UM = unmixed, M =
  mixed; ND~=~nondepleted, D = depleted. 
  Optical constant references:
  SiO$_2$,~\citet{Philipp85};
  Fe$_2$SiO$_4$,~\citet{Fabian01}, \citet{Zeidler11};
  Mg$_2$SiO$_4$,~\citet{Semenov03};
  Si,~\citet{Piller85};
  Fe,~~\citet{Semenov03};
  FeS,~\citet{Semenov03};
  Mg,~\citet{Lynch98};
  MgO,~\citet{Roessler91};
  Al$_2$O$_3$,~\citet{Toon76};
  C,~\citet{Zubko96}.
}
\tablenotetext{a}{Trace amounts of carbon dust}
\end{deluxetable}
\end{turnpage}
\clearpage

\global\pdfpageattr\expandafter{\the\pdfpageattr/Rotate 90}

\begin{thebibliography}{104}
\expandafter\ifx\csname natexlab\endcsname\relax\def\natexlab#1{#1}\fi

\bibitem[{{Abel} {et~al.}(2002){Abel}, {Bryan}, \& {Norman}}]{Abel02}
{Abel}, T., {Bryan}, G.~L., \& {Norman}, M.~L. 2002, Science, 295, 93

\bibitem[{{Asplund} {et~al.}(2009){Asplund}, {Grevesse}, {Sauval}, \&
  {Scott}}]{Asplund09}
{Asplund}, M., {Grevesse}, N., {Sauval}, A.~J., \& {Scott}, P. 2009, \araa, 47,
  481

\bibitem[{{Beers} \& {Christlieb}(2005)}]{Beers05}
{Beers}, T.~C., \& {Christlieb}, N. 2005, \araa, 43, 531

\bibitem[{{Bessell} {et~al.}(2004){Bessell}, {Christlieb}, \&
  {Gustafsson}}]{Bessell04}
{Bessell}, M.~S., {Christlieb}, N., \& {Gustafsson}, B. 2004, \apjl, 612, L61

\bibitem[{{Bianchi} \& {Schneider}(2007)}]{Bianchi07}
{Bianchi}, S., \& {Schneider}, R. 2007, \mnras, 378, 973

\bibitem[{{Bromm}(2013)}]{Bromm13}
{Bromm}, V. 2013, Reports on Progress in Physics, 76, 112901

\bibitem[{{Bromm} {et~al.}(2002){Bromm}, {Coppi}, \& {Larson}}]{Bromm02}
{Bromm}, V., {Coppi}, P.~S., \& {Larson}, R.~B. 2002, \apj, 564, 23

\bibitem[{{Bromm} {et~al.}(2001){Bromm}, {Ferrara}, {Coppi}, \&
  {Larson}}]{Bromm01}
{Bromm}, V., {Ferrara}, A., {Coppi}, P.~S., \& {Larson}, R.~B. 2001, \mnras,
  328, 969

\bibitem[{{Bromm} \& {Loeb}(2003)}]{Bromm03}
{Bromm}, V., \& {Loeb}, A. 2003, \nat, 425, 812

\bibitem[{{Bromm} {et~al.}(2009){Bromm}, {Yoshida}, {Hernquist}, \&
  {McKee}}]{Bromm09}
{Bromm}, V., {Yoshida}, N., {Hernquist}, L., \& {McKee}, C.~F. 2009, \nat, 459,
  49

\bibitem[{{Caffau} {et~al.}(2011){Caffau}, {Bonifacio}, {Fran{\c c}ois},
  {Sbordone}, {Monaco}, {Spite}, {Spite}, {Ludwig}, {Cayrel}, {Zaggia},
  {Hammer}, {Randich}, {Molaro}, \& {Hill}}]{Caffau11}
{Caffau}, E., {Bonifacio}, P., {Fran{\c c}ois}, P.,  {et~al.} 2011, \nat, 477,
  67

\bibitem[{{Cherchneff} \& {Dwek}(2009)}]{Cherchneff09}
{Cherchneff}, I., \& {Dwek}, E. 2009, \apj, 703, 642

\bibitem[{{Cherchneff} \& {Dwek}(2010)}]{Cherchneff10}
---. 2010, \apj, 713, 1

\bibitem[{{Cherchneff} \& {Lilly}(2008)}]{Cherchneff08}
{Cherchneff}, I., \& {Lilly}, S. 2008, \apjl, 683, L123

\bibitem[{{Chiaki} {et~al.}(2013{\natexlab{a}}){Chiaki}, {Nozawa}, \&
  {Yoshida}}]{Chiaki13a}
{Chiaki}, G., {Nozawa}, T., \& {Yoshida}, N. 2013{\natexlab{a}}, \apjl, 765, L3

\bibitem[{{Chiaki} {et~al.}(2013{\natexlab{b}}){Chiaki}, {Yoshida}, \&
  {Kitayama}}]{Chiaki13b}
{Chiaki}, G., {Yoshida}, N., \& {Kitayama}, T. 2013{\natexlab{b}}, \apj, 762,
  50

\bibitem[{{Christlieb} {et~al.}(2002){Christlieb}, {Bessell}, {Beers},
  {Gustafsson}, {Korn}, {Barklem}, {Karlsson}, {Mizuno-Wiedner}, \&
  {Rossi}}]{Christlieb02}
{Christlieb}, N., {Bessell}, M.~S., {Beers}, T.~C.,  {et~al.} 2002, \nat, 419,
  904

\bibitem[{{Christlieb} {et~al.}(2004){Christlieb}, {Gustafsson}, {Korn},
  {Barklem}, {Beers}, {Bessell}, {Karlsson}, \&
  {Mizuno-Wiedner}}]{Christlieb04}
{Christlieb}, N., {Gustafsson}, B., {Korn}, A.~J.,  {et~al.} 2004, \apj, 603,
  708

\bibitem[{{Ciardi} {et~al.}(2000){Ciardi}, {Ferrara}, \& {Abel}}]{Ciardi00}
{Ciardi}, B., {Ferrara}, A., \& {Abel}, T. 2000, \apj, 533, 594

\bibitem[{{Clark} {et~al.}(2008){Clark}, {Glover}, \& {Klessen}}]{Clark08}
{Clark}, P.~C., {Glover}, S.~C.~O., \& {Klessen}, R.~S. 2008, \apj, 672, 757

\bibitem[{{Clark} {et~al.}(2011){Clark}, {Glover}, {Smith}, {Greif}, {Klessen},
  \& {Bromm}}]{Clark11}
{Clark}, P.~C., {Glover}, S.~C.~O., {Smith}, R.~J.,  {et~al.} 2011, Science,
  331, 1040

\bibitem[{{Cohen} {et~al.}(2008){Cohen}, {Christlieb}, {McWilliam}, {Shectman},
  {Thompson}, {Melendez}, {Wisotzki}, \& {Reimers}}]{Cohen08}
{Cohen}, J.~G., {Christlieb}, N., {McWilliam}, A.,  {et~al.} 2008, \apj, 672,
  320

\bibitem[{{Collet} {et~al.}(2006){Collet}, {Asplund}, \&
  {Trampedach}}]{Collet06}
{Collet}, R., {Asplund}, M., \& {Trampedach}, R. 2006, \apjl, 644, L121

\bibitem[{{Cooke} {et~al.}(2011){Cooke}, {Pettini}, {Steidel}, {Rudie}, \&
  {Nissen}}]{Cooke11}
{Cooke}, R., {Pettini}, M., {Steidel}, C.~C., {Rudie}, G.~C., \& {Nissen},
  P.~E. 2011, \mnras, 417, 1534

\bibitem[{{Couchman} \& {Rees}(1986)}]{Couchman86}
{Couchman}, H.~M.~P., \& {Rees}, M.~J. 1986, \mnras, 221, 53

\bibitem[{{Donn} \& {Nuth}(1985)}]{Donn85}
{Donn}, B., \& {Nuth}, J.~A. 1985, \apj, 288, 187

\bibitem[{{Dopcke} {et~al.}(2013){Dopcke}, {Glover}, {Clark}, \&
  {Klessen}}]{Dopcke13}
{Dopcke}, G., {Glover}, S.~C.~O., {Clark}, P.~C., \& {Klessen}, R.~S. 2013,
  \apj, 766, 103

\bibitem[{{Draine}(2011)}]{DraineBook}
{Draine}, B.~T. 2011, {Physics of the Interstellar and Intergalactic Medium}
  (Princeton University Press)

\bibitem[{{Draine} \& {Li}(2001)}]{Draine01}
{Draine}, B.~T., \& {Li}, A. 2001, \apj, 551, 807

\bibitem[{{Fabian} {et~al.}(2001){Fabian}, {Henning}, {J{\"a}ger}, {Mutschke},
  {Dorschner}, \& {Wehrhan}}]{Fabian01}
{Fabian}, D., {Henning}, T., {J{\"a}ger}, C.,  {et~al.} 2001, \aap, 378, 228

\bibitem[{{Frebel}(2010)}]{Frebel10}
{Frebel}, A. 2010, Astronomische Nachrichten, 331, 474

\bibitem[{{Frebel} {et~al.}(2005){Frebel}, {Aoki}, {Christlieb}, {Ando},
  {Asplund}, {Barklem}, {Beers}, {Eriksson}, {Fechner}, {Fujimoto}, {Honda},
  {Kajino}, {Minezaki}, {Nomoto}, {Norris}, {Ryan}, {Takada-Hidai},
  {Tsangarides}, \& {Yoshii}}]{Frebel05}
{Frebel}, A., {Aoki}, W., {Christlieb}, N.,  {et~al.} 2005, \nat, 434, 871

\bibitem[{{Frebel} {et~al.}(2006){Frebel}, {Christlieb}, {Norris}, {Aoki}, \&
  {Asplund}}]{Frebel06}
{Frebel}, A., {Christlieb}, N., {Norris}, J.~E., {Aoki}, W., \& {Asplund}, M.
  2006, \apjl, 638, L17

\bibitem[{{Frebel} {et~al.}(2008){Frebel}, {Collet}, {Eriksson}, {Christlieb},
  \& {Aoki}}]{Frebel08}
{Frebel}, A., {Collet}, R., {Eriksson}, K., {Christlieb}, N., \& {Aoki}, W.
  2008, \apj, 684, 588

\bibitem[{{Frebel} {et~al.}(2007){Frebel}, {Johnson}, \& {Bromm}}]{Frebel07}
{Frebel}, A., {Johnson}, J.~L., \& {Bromm}, V. 2007, \mnras, 380, L40

\bibitem[{{Frebel} \& {Norris}(2013)}]{Frebel13a}
{Frebel}, A., \& {Norris}, J.~E. 2013, {Metal-Poor Stars and the Chemical
  Enrichment of the Universe}, ed. T.~D. {Oswalt} \& G.~{Gilmore}, 55

\bibitem[{{Fumagalli} {et~al.}(2011){Fumagalli}, {O'Meara}, \&
  {Prochaska}}]{Fumagalli11}
{Fumagalli}, M., {O'Meara}, J.~M., \& {Prochaska}, J.~X. 2011, Science, 334,
  1245

\bibitem[{{Gall} {et~al.}(2011){Gall}, {Hjorth}, \& {Andersen}}]{Gall11}
{Gall}, C., {Hjorth}, J., \& {Andersen}, A.~C. 2011, \aapr, 19, 43

\bibitem[{{Glover}(2013)}]{Glover13}
{Glover}, S. 2013, in Astrophysics and Space Science Library, Vol. 396,
  Astrophysics and Space Science Library, ed. T.~{Wiklind}, B.~{Mobasher}, \&
  V.~{Bromm}, 103

\bibitem[{{Greif} {et~al.}(2010){Greif}, {Glover}, {Bromm}, \&
  {Klessen}}]{Greif10}
{Greif}, T.~H., {Glover}, S.~C.~O., {Bromm}, V., \& {Klessen}, R.~S. 2010,
  \apj, 716, 510

\bibitem[{{Greif} {et~al.}(2008){Greif}, {Johnson}, {Klessen}, \&
  {Bromm}}]{Greif08}
{Greif}, T.~H., {Johnson}, J.~L., {Klessen}, R.~S., \& {Bromm}, V. 2008,
  \mnras, 387, 1021

\bibitem[{{Greif} {et~al.}(2011){Greif}, {Springel}, {White}, {Glover},
  {Clark}, {Smith}, {Klessen}, \& {Bromm}}]{Greif11}
{Greif}, T.~H., {Springel}, V., {White}, S.~D.~M.,  {et~al.} 2011, \apj, 737,
  75

\bibitem[{{Haiman} {et~al.}(1996){Haiman}, {Thoul}, \& {Loeb}}]{Haiman96}
{Haiman}, Z., {Thoul}, A.~A., \& {Loeb}, A. 1996, \apj, 464, 523

\bibitem[{{Hirashita} \& {Omukai}(2009)}]{Hirashita09}
{Hirashita}, H., \& {Omukai}, K. 2009, \mnras, 399, 1795

\bibitem[{{Hollenbach} \& {McKee}(1979)}]{Hollenbach79}
{Hollenbach}, D., \& {McKee}, C.~F. 1979, \apjs, 41, 555

\bibitem[{{Hosokawa} {et~al.}(2011){Hosokawa}, {Omukai}, {Yoshida}, \&
  {Yorke}}]{Hosokawa11}
{Hosokawa}, T., {Omukai}, K., {Yoshida}, N., \& {Yorke}, H.~W. 2011, Science,
  334, 1250

\bibitem[{{Jappsen} {et~al.}(2009{\natexlab{a}}){Jappsen}, {Klessen}, {Glover},
  \& {Mac Low}}]{Jappsen09a}
{Jappsen}, A.-K., {Klessen}, R.~S., {Glover}, S.~C.~O., \& {Mac Low}, M.-M.
  2009{\natexlab{a}}, \apj, 696, 1065

\bibitem[{{Jappsen} {et~al.}(2009{\natexlab{b}}){Jappsen}, {Mac Low}, {Glover},
  {Klessen}, \& {Kitsionas}}]{Jappsen09b}
{Jappsen}, A.-K., {Mac Low}, M.-M., {Glover}, S.~C.~O., {Klessen}, R.~S., \&
  {Kitsionas}, S. 2009{\natexlab{b}}, \apj, 694, 1161

\bibitem[{{Karlsson} {et~al.}(2013){Karlsson}, {Bromm}, \&
  {Bland-Hawthorn}}]{Karlsson13}
{Karlsson}, T., {Bromm}, V., \& {Bland-Hawthorn}, J. 2013, Reviews of Modern
  Physics, 85, 809

\bibitem[{{Klessen} {et~al.}(2012){Klessen}, {Glover}, \& {Clark}}]{Klessen12}
{Klessen}, R.~S., {Glover}, S.~C.~O., \& {Clark}, P.~C. 2012, \mnras, 421, 3217

\bibitem[{{Kulkarni} {et~al.}(2013){Kulkarni}, {Rollinde}, {Hennawi}, \&
  {Vangioni}}]{Kulkarni13}
{Kulkarni}, G., {Rollinde}, E., {Hennawi}, J.~F., \& {Vangioni}, E. 2013,
  arXiv:1301.4201

\bibitem[{{Larson}(1969)}]{Larson69}
{Larson}, R.~B. 1969, \mnras, 145, 271

\bibitem[{{Lynch} \& {Hunter}(1998)}]{Lynch98}
{Lynch}, D.~W., \& {Hunter}, W.~R. 1998, Handbook of Optical Constants of
  Solids III, ed. E.~D. {Palik} (Academic Press, New York), 233

\bibitem[{{Mackey} {et~al.}(2003){Mackey}, {Bromm}, \& {Hernquist}}]{Mackey03}
{Mackey}, J., {Bromm}, V., \& {Hernquist}, L. 2003, \apj, 586, 1

\bibitem[{{Mayer} \& {Duschl}(2005)}]{Mayer05}
{Mayer}, M., \& {Duschl}, W.~J. 2005, \mnras, 358, 614

\bibitem[{{Nagakura} {et~al.}(2009){Nagakura}, {Hosokawa}, \&
  {Omukai}}]{Nagakura09}
{Nagakura}, T., {Hosokawa}, T., \& {Omukai}, K. 2009, \mnras, 399, 2183

\bibitem[{{Norris} {et~al.}(2012){Norris}, {Christlieb}, {Bessell}, {Asplund},
  {Eriksson}, \& {Korn}}]{Norris12}
{Norris}, J.~E., {Christlieb}, N., {Bessell}, M.~S.,  {et~al.} 2012, \apj, 753,
  150

\bibitem[{{Norris} {et~al.}(2007){Norris}, {Christlieb}, {Korn}, {Eriksson},
  {Bessell}, {Beers}, {Wisotzki}, \& {Reimers}}]{Norris07}
{Norris}, J.~E., {Christlieb}, N., {Korn}, A.~J.,  {et~al.} 2007, \apj, 670,
  774

\bibitem[{{Norris} {et~al.}(2013){Norris}, {Yong}, {Bessell}, {Christlieb},
  {Asplund}, {Gilmore}, {Wyse}, {Beers}, {Barklem}, {Frebel}, \&
  {Ryan}}]{Norris13}
{Norris}, J.~E., {Yong}, D., {Bessell}, M.~S.,  {et~al.} 2013, \apj, 762, 28

\bibitem[{{Nozawa} \& {Kozasa}(2013)}]{Nozawa13}
{Nozawa}, T., \& {Kozasa}, T. 2013, \apj, 776, 24

\bibitem[{{Nozawa} {et~al.}(2007){Nozawa}, {Kozasa}, {Habe}, {Dwek}, {Umeda},
  {Tominaga}, {Maeda}, \& {Nomoto}}]{Nozawa07}
{Nozawa}, T., {Kozasa}, T., {Habe}, A.,  {et~al.} 2007, \apj, 666, 955

\bibitem[{{Nozawa} {et~al.}(2003){Nozawa}, {Kozasa}, {Umeda}, {Maeda}, \&
  {Nomoto}}]{Nozawa03}
{Nozawa}, T., {Kozasa}, T., {Umeda}, H., {Maeda}, K., \& {Nomoto}, K. 2003,
  \apj, 598, 785

\bibitem[{{Oh} \& {Haiman}(2002)}]{Oh02}
{Oh}, S.~P., \& {Haiman}, Z. 2002, \apj, 569, 558

\bibitem[{{Omukai}(2000)}]{Omukai00}
{Omukai}, K. 2000, \apj, 534, 809

\bibitem[{{Omukai} {et~al.}(2010){Omukai}, {Hosokawa}, \& {Yoshida}}]{Omukai10}
{Omukai}, K., {Hosokawa}, T., \& {Yoshida}, N. 2010, \apj, 722, 1793

\bibitem[{{Omukai} {et~al.}(2005){Omukai}, {Tsuribe}, {Schneider}, \&
  {Ferrara}}]{Omukai05}
{Omukai}, K., {Tsuribe}, T., {Schneider}, R., \& {Ferrara}, A. 2005, \apj, 626,
  627

\bibitem[{{Paquette} \& {Nuth}(2011)}]{Paquette11}
{Paquette}, J.~A., \& {Nuth}, III, J.~A. 2011, \apjl, 737, L6

\bibitem[{Patnaik(2003)}]{HICC}
Patnaik, P. 2003, Handbook of Inorganic Chemical Compounds, McGraw-Hill
  Handbooks (McGraw-Hill Professional Publishing)

\bibitem[{{Philipp}(1985)}]{Philipp85}
{Philipp}, H.~R. 1985, Handbook of Optical Constants of Solids, ed. E.~D.
  {Palik} (Academic Press, New York), 719

\bibitem[{{Piller}(1985)}]{Piller85}
{Piller}, H. 1985, Handbook of Optical Constants of Solids, ed. E.~D. {Palik}
  (Academic Press, New York), 571

\bibitem[{{Pollack} {et~al.}(1994){Pollack}, {Hollenbach}, {Beckwith},
  {Simonelli}, {Roush}, \& {Fong}}]{Pollack94}
{Pollack}, J.~B., {Hollenbach}, D., {Beckwith}, S.,  {et~al.} 1994, \apj, 421,
  615

\bibitem[{{Roessler} \& {Hunter}(1991)}]{Roessler91}
{Roessler}, D.~M., \& {Hunter}, D.~R. 1991, Handbook of Optical Constants of
  Solids II, ed. E.~D. {Palik} (Academic Press, New York), 919

\bibitem[{{Safranek-Shrader} {et~al.}(2012){Safranek-Shrader}, {Agarwal},
  {Federrath}, {Dubey}, {Milosavljevi{\'c}}, \& {Bromm}}]{SafShrad12}
{Safranek-Shrader}, C., {Agarwal}, M., {Federrath}, C.,  {et~al.} 2012, \mnras,
  426, 1159

\bibitem[{{Safranek-Shrader} {et~al.}(2010){Safranek-Shrader}, {Bromm}, \&
  {Milosavljevi{\'c}}}]{SafShrad10}
{Safranek-Shrader}, C., {Bromm}, V., \& {Milosavljevi{\'c}}, M. 2010, \apj,
  723, 1568

\bibitem[{{Safranek-Shrader} {et~al.}(2013){Safranek-Shrader}, {Milosavljevic},
  \& {Bromm}}]{SafShrad13}
{Safranek-Shrader}, C., {Milosavljevic}, M., \& {Bromm}, V. 2013,
  arXiv:1307.1982

\bibitem[{Salvaterra {et~al.}(2004)Salvaterra, Ferrara, \&
  Schneider}]{Salvaterra04}
Salvaterra, R., Ferrara, A., \& Schneider, R. 2004, New Astronomy, 10, 113

\bibitem[{{Santoro} \& {Shull}(2006)}]{Santoro06}
{Santoro}, F., \& {Shull}, J.~M. 2006, \apj, 643, 26

\bibitem[{{Schneider} {et~al.}(2002){Schneider}, {Ferrara}, {Natarajan}, \&
  {Omukai}}]{Schneider02}
{Schneider}, R., {Ferrara}, A., {Natarajan}, P., \& {Omukai}, K. 2002, \apj,
  571, 30

\bibitem[{{Schneider} {et~al.}(2004){Schneider}, {Ferrara}, \&
  {Salvaterra}}]{Schneider04}
{Schneider}, R., {Ferrara}, A., \& {Salvaterra}, R. 2004, \mnras, 351, 1379

\bibitem[{{Schneider} \& {Omukai}(2010)}]{Schneider10}
{Schneider}, R., \& {Omukai}, K. 2010, \mnras, 402, 429

\bibitem[{{Schneider} {et~al.}(2012{\natexlab{a}}){Schneider}, {Omukai},
  {Bianchi}, \& {Valiante}}]{Schneider12}
{Schneider}, R., {Omukai}, K., {Bianchi}, S., \& {Valiante}, R.
  2012{\natexlab{a}}, \mnras, 419, 1566

\bibitem[{{Schneider} {et~al.}(2006){Schneider}, {Omukai}, {Inoue}, \&
  {Ferrara}}]{Schneider06}
{Schneider}, R., {Omukai}, K., {Inoue}, A.~K., \& {Ferrara}, A. 2006, \mnras,
  369, 1437

\bibitem[{{Schneider} {et~al.}(2012{\natexlab{b}}){Schneider}, {Omukai},
  {Limongi}, {Ferrara}, {Salvaterra}, {Chieffi}, \& {Bianchi}}]{Schneider12b}
{Schneider}, R., {Omukai}, K., {Limongi}, M.,  {et~al.} 2012{\natexlab{b}},
  \mnras, 423, L60

\bibitem[{{Semenov} {et~al.}(2003){Semenov}, {Henning}, {Helling}, {Ilgner}, \&
  {Sedlmayr}}]{Semenov03}
{Semenov}, D., {Henning}, T., {Helling}, C., {Ilgner}, M., \& {Sedlmayr}, E.
  2003, \aap, 410, 611

\bibitem[{{Shi} {et~al.}(2009){Shi}, {Gehren}, {Mashonkina}, \& {Zhao}}]{Shi09}
{Shi}, J.~R., {Gehren}, T., {Mashonkina}, L., \& {Zhao}, G. 2009, \aap, 503,
  533

\bibitem[{{Silvia} {et~al.}(2010){Silvia}, {Smith}, \& {Shull}}]{Silvia10}
{Silvia}, D.~W., {Smith}, B.~D., \& {Shull}, J.~M. 2010, \apj, 715, 1575

\bibitem[{{Simcoe} {et~al.}(2012){Simcoe}, {Sullivan}, {Cooksey}, {Kao},
  {Matejek}, \& {Burgasser}}]{Simcoe12}
{Simcoe}, R.~A., {Sullivan}, P.~W., {Cooksey}, K.~L.,  {et~al.} 2012, \nat,
  492, 79

\bibitem[{{Smith} {et~al.}(2009){Smith}, {Turk}, {Sigurdsson}, {O'Shea}, \&
  {Norman}}]{Smith09}
{Smith}, B.~D., {Turk}, M.~J., {Sigurdsson}, S., {O'Shea}, B.~W., \& {Norman},
  M.~L. 2009, \apj, 691, 441

\bibitem[{{Stacy} {et~al.}(2010){Stacy}, {Greif}, \& {Bromm}}]{Stacy10}
{Stacy}, A., {Greif}, T.~H., \& {Bromm}, V. 2010, \mnras, 403, 45

\bibitem[{{Stacy} {et~al.}(2012){Stacy}, {Greif}, \& {Bromm}}]{Stacy12}
---. 2012, \mnras, 422, 290

\bibitem[{{Suda} {et~al.}(2008){Suda}, {Katsuta}, {Yamada}, {Suwa}, {Ishizuka},
  {Komiya}, {Sorai}, {Aikawa}, \& {Fujimoto}}]{Suda08}
{Suda}, T., {Katsuta}, Y., {Yamada}, S.,  {et~al.} 2008, \pasj, 60, 1159

\bibitem[{{Tegmark} {et~al.}(1997){Tegmark}, {Silk}, {Rees}, {Blanchard},
  {Abel}, \& {Palla}}]{Tegmark97}
{Tegmark}, M., {Silk}, J., {Rees}, M.~J.,  {et~al.} 1997, \apj, 474, 1

\bibitem[{{Todini} \& {Ferrara}(2001)}]{Todini01}
{Todini}, P., \& {Ferrara}, A. 2001, \mnras, 325, 726

\bibitem[{{Tohline}(1980)}]{Tohline80}
{Tohline}, J.~E. 1980, \apj, 239, 417

\bibitem[{{Toon} {et~al.}(1976){Toon}, {Pollack}, \& {Khare}}]{Toon76}
{Toon}, O.~B., {Pollack}, J.~B., \& {Khare}, B.~N. 1976, \jgr, 81, 5733

\bibitem[{{Tsuribe} \& {Omukai}(2006)}]{Tsuribe06}
{Tsuribe}, T., \& {Omukai}, K. 2006, \apjl, 642, L61

\bibitem[{{Tumlinson}(2006)}]{Tumlinson06}
{Tumlinson}, J. 2006, \apj, 641, 1

\bibitem[{{Whalen} {et~al.}(2008){Whalen}, {van Veelen}, {O'Shea}, \&
  {Norman}}]{Whalen08}
{Whalen}, D., {van Veelen}, B., {O'Shea}, B.~W., \& {Norman}, M.~L. 2008, \apj,
  682, 49

\bibitem[{{Wise} \& {Abel}(2007)}]{Wise07}
{Wise}, J.~H., \& {Abel}, T. 2007, \apj, 665, 899

\bibitem[{{Yong} {et~al.}(2013){Yong}, {Norris}, {Bessell}, {Christlieb},
  {Asplund}, {Beers}, {Barklem}, {Frebel}, \& {Ryan}}]{Yong13}
{Yong}, D., {Norris}, J.~E., {Bessell}, M.~S.,  {et~al.} 2013, \apj, 762, 26

\bibitem[{{Yoshida} {et~al.}(2003){Yoshida}, {Abel}, {Hernquist}, \&
  {Sugiyama}}]{Yoshida03}
{Yoshida}, N., {Abel}, T., {Hernquist}, L., \& {Sugiyama}, N. 2003, \apj, 592,
  645

\bibitem[{{Zeidler} {et~al.}(2011){Zeidler}, {Posch}, {Mutschke}, {Richter}, \&
  {Wehrhan}}]{Zeidler11}
{Zeidler}, S., {Posch}, T., {Mutschke}, H., {Richter}, H., \& {Wehrhan}, O.
  2011, \aap, 526, A68

\bibitem[{{Zhang} {et~al.}(2011){Zhang}, {Karlsson}, {Christlieb}, {Korn},
  {Barklem}, \& {Zhao}}]{Zhang11}
{Zhang}, L., {Karlsson}, T., {Christlieb}, N.,  {et~al.} 2011, \aap, 528, A92

\bibitem[{{Zubko} {et~al.}(1996){Zubko}, {Mennella}, {Colangeli}, \&
  {Bussoletti}}]{Zubko96}
{Zubko}, V.~G., {Mennella}, V., {Colangeli}, L., \& {Bussoletti}, E. 1996,
  \mnras, 282, 1321

\end{thebibliography}
\end{document}